\documentclass[12pt,preprint]{aastex}

\usepackage{mathtools}
\usepackage{amstext}
\usepackage{natbib}

\usepackage{color}

\usepackage{textcomp}
\usepackage[normalem]{ulem}
\usepackage{enumerate}

\def\actaa{\ref@jnl{Acta Astron.}}      

\pdfoutput=1

\newcommand{\pd}{\partial}
\newcommand{\Msy}{M_\odot\, \rm{yr}^{-1}}

\newcommand{\MdotAccr}{\dot{M}_{{\rm a}}}

\makeatother

\begin{document}
\author{A. Dorodnitsyn\altaffilmark{1,2,3}, T. Kallman\altaffilmark{1}}
\altaffiltext{1}{Laboratory for High Energy Astrophysics, NASA Goddard Space 
Flight Center, Code 662, Greenbelt, 
MD, 20771, USA}
\altaffiltext{2}{University of Maryland, Baltimore County (UMBC/CRESST), Baltimore, MD 21250, USA}
\altaffiltext{3}{Space Research Institute, 84/32, Profsoyuznaya st., Moscow, Russia}

\title{
Parsec-scale obscuring accretion disk with large scale magnetic field in  AGN
}
\begin{abstract}
Magnetic field dragged from the galactic disk along with inflowing gas 
can provide vertical support to the geometrically and optically thick 
pc-scale torus in AGN.  Using the Soloviev solution initially  developed for 
Tokamaks 
we derive an analytical model for a rotating torus supported and confined by 
magnetic field.
We further perform three-dimensional magneto-hydrodynamics
simulations of X-ray irradiated pc-scale magnetized tori.  We follow the time 
evolution and compare models which adopt initial conditions derived from our 
analytic model with simulations in which the initial magnetic flux is entirely contained 
within the gas torus.
Numerical simulations demonstrate that the initial conditions based on
the analytic solution produce a longer-lived torus and one which produces 
obscuration which is generally consistent with observed constraints.
\end{abstract}

\section{Introduction}

The unification paradigm of Active Galactic Nuclei
(AGN) implies the existence of a geometrically thick belt of matter
which wraps and hides the inner, most luminous region of AGN  from viewing close
to the equatorial plane
\citep{Antonucci84,AntonucciMiller1985,Rowan-Robinson77}.  This
can explain the dichotomy of
Type I and Type II AGNs and Quasars, attributing their differing appearance
to the geometrical 
positioning of an
AGN with respect to an observer. 

Observations  favor this model,
but  the mechanism that actually supports such a geometrically thick structure 
is not determined.
In the absence of efficient cooling, the virial theorem predicts that
gas which is orbiting the black hole at the radius of the putative torus,
$r\simeq 1\,{\rm pc}$
will have temperature $T_{{\rm vir,g}}\simeq10^{6}$ K 
for a $10^7 M_\odot$ black hole.
Dust at such temperatures is destroyed by sputtering, while the presence of dust 
in the torus is 
supported by abundant observational evidence.
Emission from warm dust manifests itself in a broad hump in the spectral energy 
distribution (SED) in the
$\sim 1 - 10 \mu$ wavelength band 
\citep{Rieke81,Barvainis87,Sanders89,PierKrolik93}.
Interferometric observations provide direct evidence of warm,
multi-temperature dust at parsec scales in a growing sample of nearby type II 
and type I AGNs,  including
NGC 1068 \citep{Poncelet06,Raban09,Wittkowski2004,Jaffe2004} the Circinus galaxy
\citep{Tristram07,Tristram14} and others 
\citep{Kishimoto13,Honig12,Meisenheimer07,Beckert08}.
Observations of ionization and scattering cones  also indicate the existence of 
a toroidal obscuring 
structure \citep[eg.][]{Pogge93,Wilson96,Zakamska05}.

Obscuration could be provided by a globally warped but intrinsically 
geometrically thin disk \citep{Phinney89,Sanders89}. A disk that is severely 
warped at parsec scales may also be the result of stochastic accretion, i.e. 
accretion that proceeds from random directions \citep{Lawrence2010K,Hopkins12}.
Numerical simulations which support this scenario, and follow AGN fueling from 
galactic scales \citep[eg.][]{Hopkins05,Hopkins10,Hopkins12}, are very demanding 
and thus are somewhat restricted in the ways they take into account effects of magnetic fields or radiation transfer.
When they provide geometrically thick inflows the gas temperature always resides at a 
significant fraction of $T_{{\rm vir,g}}$.
UV and IR radiation pressure on dust may provide the necessary vertical support.
The temperature of such a torus will be of the order of the "radiation virial 
temperature''
$T_{\rm vir,r} \simeq 10^3\, \left( n_7 \, M_7/ r_{\rm pc} \right) ^{1/4} \,{\rm K} \ll T_{\rm vir,g}$, where
$n = 10^7\,n_7 \, {\rm cm}^{-3}$ is the number density and  $M_{\rm BH} =10^7\,M_7 M_\odot$ is the mass of the Black Hole, and  $r_{\rm pc}$ is the radius in parsecs.

Several such models have been proposed: a static torus: 
\citep[][]{Krolik07,ShiKrolik08} 
as well as models in which the torus is dynamic - ``windy torus'':
\citep[][]{Dorodnitsyn11a}, 
\citep[][]{Dorodnitsyn12a,Dorodnitsyn12b,ChanKrolik16}. 
Simulations show that IR pressure support works most efficiently in AGNs
with inferred bolometric luminosity $L\gtrsim0.1\,L_{{\rm edd}}$,
where $L_{{\rm edd}}$ is the Eddington luminosity.

The  torus is also the most likely source of gas for the inner
accretion disk and the fueling of the black hole.
This raises the important question of the nature of the mechanism that provides
angular momentum redistribution.  This includes the mechanism which operates on 
scales larger than the torus,
and brings gas into the torus, and also the mechanisms operating within the 
torus.

On galactic scales, gas can be funneled towards the center as a result
of a galaxy merger, leading to the development of the bar instability
\citep[ ][]{Shlosman89,Shlosman90}.  Observations indicate the
existence of large scale magnetic fields on galactic scales
\citep[i.e.][]{Beck2013, Beck11}, and the inflow can transport
magnetic flux from the inner parts of the host galaxy towards the
torus. The importance of magnetic flux can be inferred from the following 
approximate arguments: 
assume the spherically-symmetric shell of gas located within the BH sphere-of-influence, falls radially at roughly 
a free-fall velocity and with a constant accretion rate. Then
assuming perfect conductivity, conservation of magnetic flux
implies $B_{r}\sim r^{-2}$, where $B_r$ is the radial component of the
magnetic field.  The gas internal energy scales as
$\rho v^{2}/2\sim r^{-5/2}$.  The density profile is then $\rho\sim r^{-3/2}$ and
the distance from the BH where magnetic field becomes dynamically
important can be estimated from energy equipartition arguments
\citep{BKRuz74,BKLovelace2000}:
$r_{m}=r_{{\rm out}}^{8/3} B_{{\rm out}}^{4/3}
\,\dot{M}_{\odot}^{-2/3} (GM_{\rm BH })^{-1/3}\simeq 81 B_{10}^{4/3}
R_{100}^{8/3} \,\dot{m}_{0.01}^{-2/3} M_{7}^{-1/3}$pc,
where $B_{10}$ is the galactic magnetic field scaled to
$10\,\mu{\rm G}$, $R_{100}$ is the outer AGN radius $r_{{\rm out}}$ in
units of 100 pc, the accretion rate, $\dot{m}_{0.01}$ is scaled to
$10^{-2}M_{\odot}/{\rm yr}$, and $M_{7}$ is the mass of the BH in
$10^{7}M_{\odot}$.  Parsec-scale magnetic field loops that were
small-scale in the galactic disk will dominate the geometry in the
torus region.  This illustrates the potential dynamical importance of
magnetic field in the torus.  If so, in such a magnetized rotating
torus, orbital shear can lead to magnetic instability in which angular
momentum is redistributed via magnetic stresses.

In AGN disks various angular momentum transport mechanisms are likely to 
operate at different scales.  
Long-range non-axisymmetric gravitational torques acting in the stellar and gas 
system
can be responsible for angular momentum transport on galactic scales 
\citep{LyndenBellKalnajs72}. Theoretical results 
\citep[][]{Shlosman89,Shlosman90} along with the results of numerical 
simulations \citep{Hopkins12} suggest that global non-axisymmetric instabilities 
are capable of siphoning significant amounts of gas towards a central AGN. A 
different regime is established if the gas cooling time is comparable to the 
dynamical time $\simeq \Omega^{-1}$.  Such gas settles into a 
"gravito-turbulent state" which is characterized by the non-linear development of 
self-gravitating instabilities which provide dissipation and stresses, eventually 
leading to angular momentum transport 
\citep[i.e.][]{Paczynski78,Gammie01,Goodman03,Rafikov09}.

In the inner accretion disk MHD turbulence driven by the MRI instability 
\citep{BalbusHawley91} is
currently the most likely mechanism for angular momentum transport leading to 
accretion 
Significant insight here comes from local, shearing box simulations. An 
important 
result of such simulations is that the presence of net magnetic flux in the 
simulation box
essentially guarantees the efficacy of the MRI.  It has been shown 
\citep{FromangPapaloizou07},
that in the absence of such net magnetic flux, the saturated level of
magnetic stresses due to MRI decreases with increasing numerical resolution.  
Further simulations in shearing boxes showed that 
preexisting net large-scale magnetic flux not only increases MRI-driven
turbulence but could be a necessary ingredient for the saturation of
MRI-stresses at a level relevant to angular transport in standard $\alpha$ disks
\citep[][]{Shi2016,Hawley95SBox,Sano04,Simon09,Guan09}.
These results imply that the magnetic flux advected from the galaxy can enhance 
turbulent transport in the torus.

Shearing box simulations cannot assess non-local coupling between distant disk 
regions.
Such non-locality can be provided by large-scale field connecting different 
patches of the disk, and
can only be addressed  via global simulations.
The potential effect of large scale fields on accretion dynamics have been 
pointed out by 
early theoretical \citep{BKRuz74} and numerical 
\citep{Igumenshchev03,Narayanetal03} studies. 
Current understanding of angular momentum transport and MHD turbulence connects 
the results from global and local simulations 
\citep[e.g.][]{Hawley2013,Sorathia12}.
One insight is the importance of accumulation of the vertical magnetic flux in 
the inner parts of the disk - an effect with a potential to significantly affect 
accretion leading to a ``magnetically arrested disk'' 
\citep[e.g.][]{Igumenshchev08,Beckwith09,Suzuki2014, 
Avara16,Beckwith11,McKinney12}

Surprisingly little effort has been devoted to the investigation of the role of  
large scale magnetic fields in the context of the AGN outer disk/torus.
\citet{Lovelace98} derived a model of a magnetically-supported,  thick
torus assuming that the magnetic field is produced by an
equatorial current loop and making a number of other
simplifications suitable for an approximate analytic model.
A different class of magnetic models
associate the torus with a geometrically thick flow, driven either by magnetic 
driving
\citep{KoniglKartje94,ElitzurShlosman06,Fukumura2010} alone, or in combination 
with radiation pressure \citep{Emmering92,Everett05,Keating12}. To launch the 
wind these models rely on an ordered global magnetic field with foot-points 
originating on a thin disk.  

To explore conditions needed to produce obscuration and outflows, and to address 
the effect of anomalous angular momentum transport \citet{Dorodnitsyn16} 
performed viscous radiation hydrodynamics simulations.  These simulations show 
that the obscuring  torus can effectively supply the inner accretion disc with 
gas, being at the same time geometrically thick due to radiation pressure of IR 
on dust and X-ray heating. An effective viscosity  was adopted to disguise 
unknown details of self-gravity  and magnetic stresses. This approach however 
cannot mimic the role of large scale magnetic field.
A complementary study \citep{ChanKrolik16} treated a weakly magnetized torus 
exposed to external UV radiation. They did not take into account X-ray heating 
and dust evaporation in the low density regions of the torus such as in the 
torus funnel. Despite these differences the results of these simulations are 
broadly consistent with the results of \citet{Dorodnitsyn11a, Dorodnitsyn12a} 
who concluded that at high AGN luminosity the torus dynamics is largely 
determined by just two parameters: radiation input and high dust opacity.

The effect of global field geometry on disk structure has been studied in
the context of formation of jets and winds \citep{Beckwith09}.  Various
assumptions about the initial field geometry and
its effects on the net accretion rate have been explored by
McKinney and Blandford (2009) and Beckwith et al. (2008).  In the context of the 
pc-scale torus, an
important question is the relation of the global field to the torus aspect ratio
and its ability to provide obscuration.  The study of this question
is primary goal of this paper.

The properties of the torus depend on the interplay between radiation,
hydrodynamics, gravity and magnetic field.
Our program so far has been to include the first two of these.
It is the goal of the current paper to test the effects of global magnetic field 
on the 
structure and evolution of the torus, by treating the torus essentially as a 
magnetized accretion disk
exposed to  external irradiation. 
In this paper we will be exploring models for the torus which include magnetic fields and 
X-rays heating but do not include any other radiation effects.

Observational evidence suggests that the distribution of hot dust in AGN scales with the square root of the luminosity \citep[i.e.]{Barvainis87,Tristram2011,Suganuma06,Kishimoto11,Weigelt12,LopezGonzaga16}
The connection of the inner torus to the BLR \citep[i.e.][]{Netzer93} was 
probed in \citep[e.g.][]{CzernyHryniewicz11, Czerny04}. In our previous research that involved radiation-hydrodynamics studies of AGN torus, dust sublimation radius appeared as a natural scale of the AGN along with the fact that complicated radiation transfer effects intervene in a significant way. In the current work we are interested mainly in the effects of the magnetic fields and no dust or radiation pressure on dust is considered.
Correspondingly, our current studies do not contain any such scale as the dust sublimation radius.

The plan of the paper is as follows:
In the first part we study an analytic model of a magnetically supported torus.
Our approximate solution is based on the exact solution of the Grad-Shafranov 
equation - the solution initially derived by Soloviev for the toroidal 
equilibrium in Tokamaks.
This part of the paper is complimentary to the second  where we perform three-
dimensional numerical
simulations of the X-ray illuminated torus threaded by global  magnetic field.
We compare simulations using initial conditions based on our analytic model with 
simulations
using initial conditions in which the initial field is entirely within the torus 
gas.
We show that the evolution of a physically realistic torus is sensitive to these 
assumptions
about the global field distribution,
and that this affects the obscuring properties as well.
In the results section we summarize many of the 
characteristics of the torus such as its dynamics, obscuring properties, 
accretion-rate etc.
We conclude with discussion of the implications of our models for understanding 
the torus dynamics, and
potential limitations associated with the our assumptions.

\section{MHD Equations for AGN Torus}

We model the torus using three-dimensional, ideal MHD equations  describing 
magnetized gas
susceptible to radiative heating and cooling in a Newtonian
gravitational potential:

\begin{eqnarray}
  \text{\ensuremath{\partial}}_{t}\rho+\text{\ensuremath{\nabla}}\cdot(\rho\:\mathbf{v}) & = & 0\mbox{,}\label{eq:contConservEq}\\
\text{\ensuremath{\partial}}_{t}(\rho\mathbf{v})+\text{\ensuremath{\nabla}}\cdot(\rho\mathbf{vv}\text{\textminus}\mathbf{BB}+P^{*}\mathbf{I}) & = & -\rho\frac{G\thinspace M_{{\rm BH}}}{r^{2}}\thinspace{\bf \hat{r}}\mbox{,}\label{eq:momentConservEq}\\
\text{\ensuremath{\partial}}_{t}E+\text{\ensuremath{\nabla}}\cdot[(E+P^{*})\mathbf{v}\text{\textminus}\mathbf{B}(\mathbf{B\text{\ensuremath{\cdot}}v})] & = & -\rho\mathbf{v}\text{\ensuremath{\cdot}}\text{\ensuremath{\frac{G\thinspace M_{{\rm BH}}}{r^{2}}}
\thinspace\ensuremath{\hat{\mathbf{r}}}} -\rho\,\cal{L},\label{eq:energyConservEq}\\ \text{\ensuremath{\partial}}_{t}\mathbf{B}+\text{\ensuremath{\nabla}}\cdot(\mathbf{v\text{\ensuremath{\cdot}}B}\text{\textminus}\mathbf{B\text{\ensuremath{\cdot}}v}) & = & 0\mbox{,}\label{eq:magnConservEq}
\end{eqnarray}
where  density, $\rho= n \, m_{\rm u} \mu$
($m_{\rm u} = 1\,\text{amu} \simeq m_p$ is the atomic mass unit;
$m_p$ is the proton's mass); we also adopt $\mu=1$ for the mean
mean molecular weight. 
$P^{*}\text{\ensuremath{\equiv}}P+(\mathbf{B \cdot B})/2$
is the total pressure, $E$ is the total energy density,
$E\equiv
e+\rho(\mathbf{v}\cdot\mathbf{v})/2+(\mathbf{B}\cdot\mathbf{B})/2$,
where $e$ is the internal energy density.  A  polytropic equation of
state is assumed: $P=(\gamma-1)\,e$, where $P$ is the gas pressure, $\gamma$ is the ratio
of specific heats.   Additional notation includes $r$, and
${\bf \hat r}$ - the usual spherical radius and the spherical
radial unit vector. Interaction with radiation is
included in the form of the X-ray cooling-heating function,
$\cal{L} \left[ {\rm erg \, g^{-1} \, s^{-1}}\right]$ in
(\ref{eq:energyConservEq}) (see Section \ref{Xrays}). In Section 
\ref{sec:Numerical-Solution}
we provide numerical solution to these equations (\ref{eq:contConservEq})--
(\ref{eq:magnConservEq}).
As a prelude,  we first discuss equilibrium solutions for a magnetic torus.

\subsection{Soloviev solution for the magnetically-confined
  torus\label{sec:ApproxSolOfTorus}}

For an illustration of the possibility of magnetic torus 
support, we notice that analytic estimates for the torus
density distribution can be made upon making several simplifying
assumptions: i) the torus is stationary, and ii) axially-symmetric.
The latter implies that the dependence on the
three components of the magnetic field can be reduced to the
dependence on a single scalar function (see the Appendix) for
which one can take a magnetic flux through the circle of a radius,
$R$:

\begin{equation}
  \Psi(R,z)=\frac{1}{2\pi}\int{\bf B}\cdot{\bf n}\:ds=
  A_{\phi}R\mbox{,}\label{eq:psiDefShort}
\end{equation}
where ${\bf n}$ is the normal vector to the equatorial plane,
$A_\phi$ is the $\phi$-component of the vector potential.  Then 
the magnetic field can  be expressed as

\begin{equation}  
\mathbf{B}=\frac{I}{R}\,\hat{\phi}+\frac{\nabla\Psi\times\hat{\phi}}{R}\mbox{,}\
\label{eq:B_vector_from_Psi}
\end{equation}
where $I=RB_{\phi}$ is the poloidal current, and $\hat{\phi}$ is
the unit vector in $\phi$ direction.

\begin{figure}
  \includegraphics[scale=0.9]{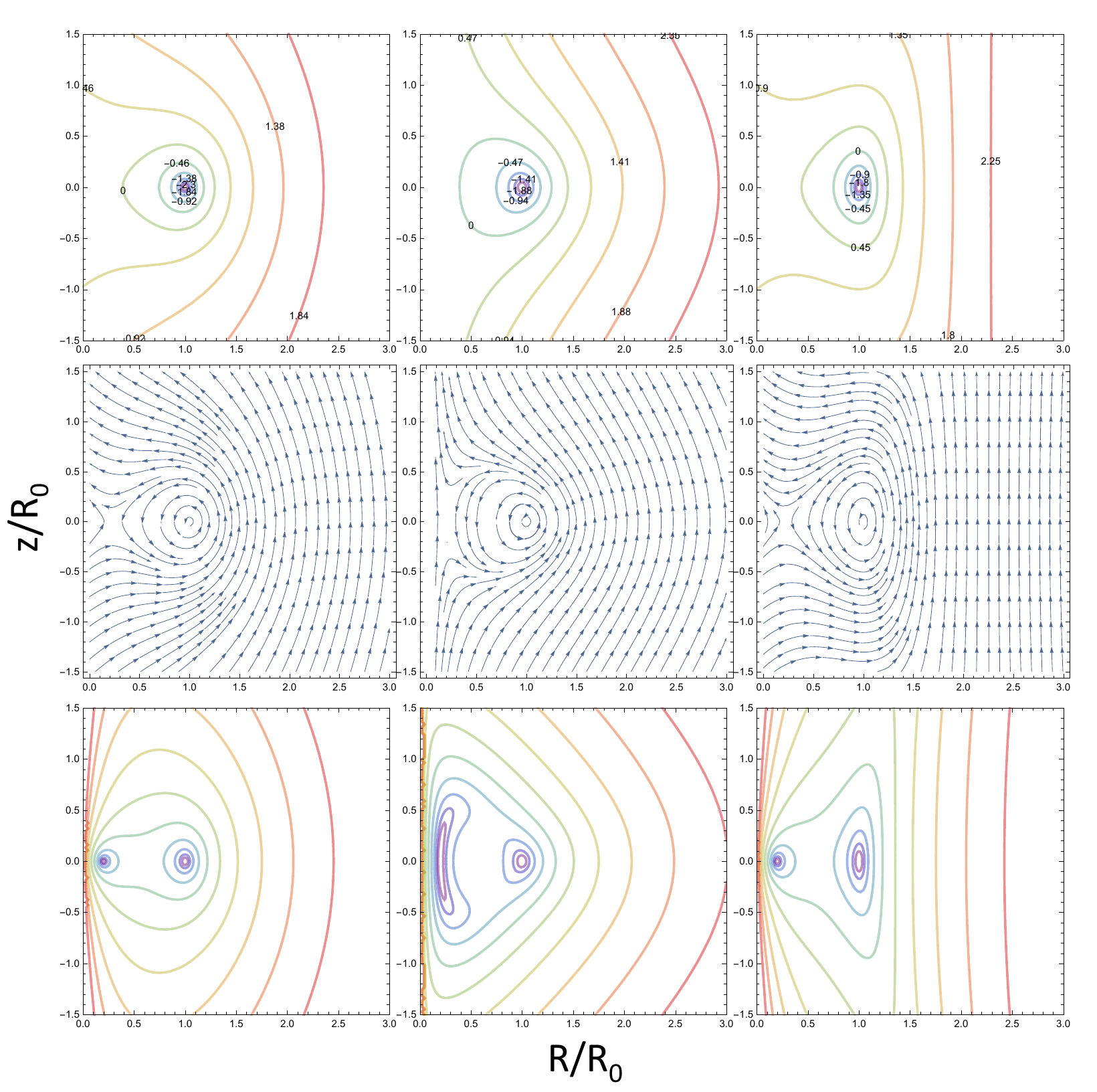}
  \caption{Flux function, $\log\Psi$ (upper), magnetic field,
    (middle), and magnetic pressure, $\log P_{{\rm m}}$ (lower)
    panel. Axes: horizontal: $R/R_0$; vertical: $z/R_0$, where $R_0$
    is the radius of the magnetic axis. Parameters $a_{1},\:a_{2},\,b_{1}$ from
    (\ref{eq:psiSolShort}) are fixed for all plots in a column: left
    column: $a_{1}=1,\:a_{2}=1,\,b_{1}=3$; middle column:
    $a_{1}=3,\:a_{2}=1,\,b_{1}=3$; right column:
    $a_{1}=-1,\:a_{2}=3,\,b_{1}=4$.}
    \label{FluxFuncPlot}
\end{figure}

In a seminal work, Soloviev \citep{Leontovich_Vol7_Soloviev} gives
an approximate solution to the problem of toroidal equilibrium
in Tokamaks.  Considering a non-rotating, magnetically confined
and gas supported torus, Soloviev solves the corresponding
Grad-Shafranov force balance equation (see the Appendix).  He
argues that the flux function, $\Psi(R,z)$ in such a case can be
expanded around the magnetic axis in a power series of
$(R^{2}-R_{0}^{2})/R_{0}^{2}$ and $z^{2}/R_{0}^{2}$, where $R_{0}$
is the cylindrical radius of the magnetic axis. On the magnetic
axis, i.e. at $R=R_{0},$ $z=0$ it should satisfy
$\partial_{R}\Psi=\partial_{z}\Psi=0$. An additional condition is
provided by the requirement that $\Psi$ should be symmetrical
with respect to the equatorial plane. The resultant solution to
the Grad-Shafranov equations can be cast in the form:

\begin{equation}
  \Psi(R,z)=z^{2}\left(a_{1}\left(R^{2}-
R_{0}^{2}\right)+b_{1}\right)+a_{2}\left(R^{2}-
R_{0}^{2}\right)^{2}\mbox{{,}}\label{eq:psiSolShort}
\end{equation}

\noindent where $a_{1},\:a_{2,}$and $b_{1}$ are parameters.  
Figure \ref{FluxFuncPlot} shows contour plots of $\log \Psi$,  poloidal field, 
${\bf B}_\text{p}$, and $\log P_\text{m}$ where $P_\text{m}=B^2/8\pi$ is the 
magnetic pressure.
A class of
solutions was found by Soloviev, assuming that, $\frac{dP}{d\Psi}=Const$.
If so, parameters $a_{1}, \: a_{2},$ and $b_{1}$ are found upon
substitution to the Grad-Shafranov equation
\citep{Leontovich_Vol7_Soloviev}.  Rotation was included into the Soloviev-type 
solution by 
\cite{MaschkePerrin80}; To include rotation in this scheme it was found that it 
is necessary to adopt a significant number of further additional 
simplifications. The geometry of the $\Psi=const.$ surfaces can be complex:
Depending on the parameters, a second magnetic axis can be found at the origin of the coordinate system; hyperbolic points can occur at the intersection of the separatrices, etc.  
These are important for the stability of the Tokamak, although in the case of an accretion disk such fine details of the initial distribution of the magnetic field are quickly ``forgotten'' by the turbulent solution.

\begin{figure}
  \includegraphics[scale=1]{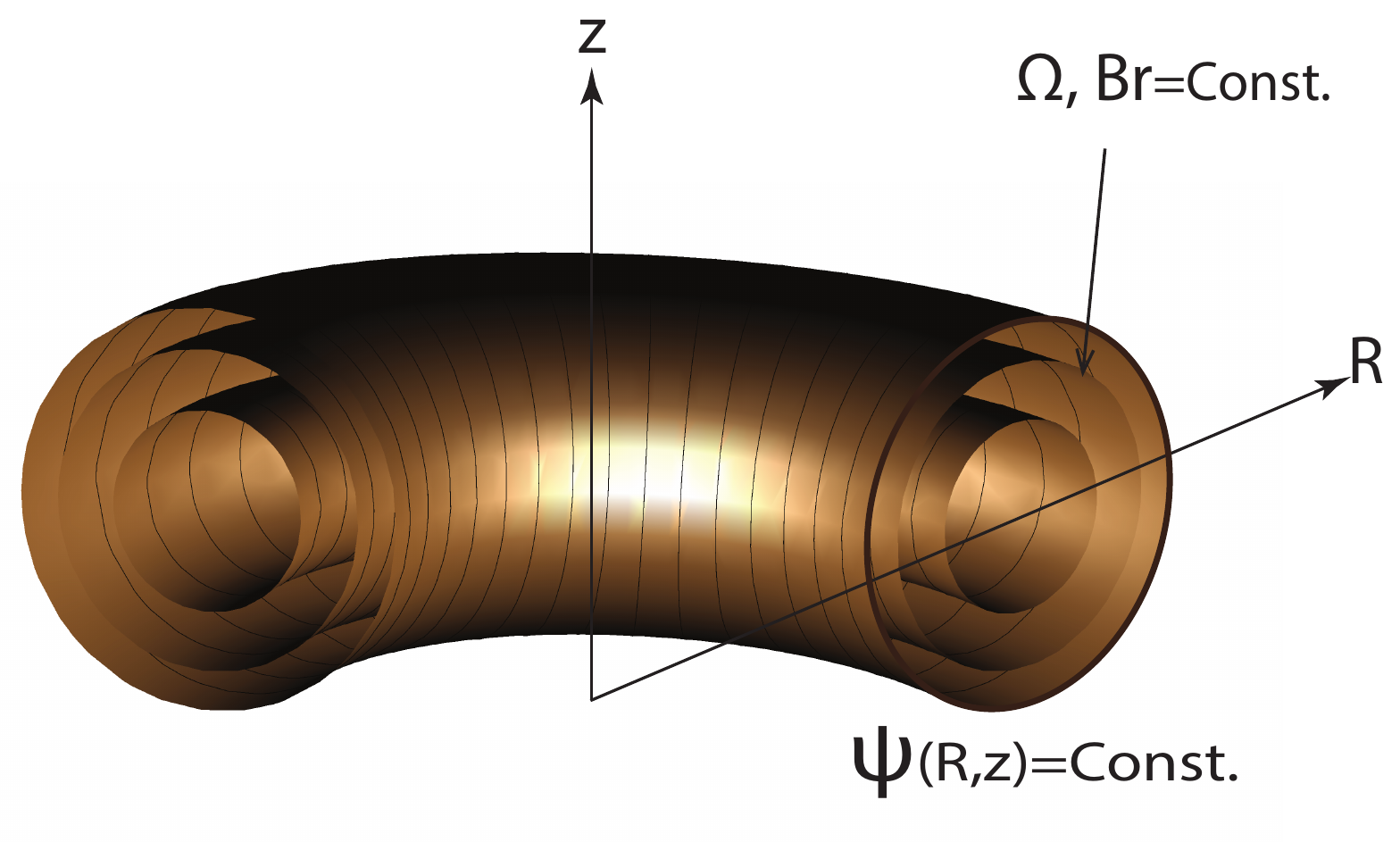}\caption{A sketch of a
    segment of the torus composed of the of nested magnetic flux
    surfaces, $\Psi$ from (\ref{eq:psiDefShort}). Gas that belongs
    to a particular flux surface has a fixed value of angular
    velocity, $\Omega(\Psi)$ and Bernoulli function,
    $Br(\Psi)$ from (\ref{eq:BrFunc1}). 
    Thin lines mark the poloidal magnetic field, ${\bf B}_p$.}
    \label{TorusSketchFluxSurf}
\end{figure}

In this work we take a different approach.  From the theory of magnetically-
driven winds it is known that
moving gas in the magnetic field should conform to certain integrals
of motion.
For example, the equation of motion can be projected onto the direction
of the poloidal magnetic field, ${\bf B}_{p}$ to obtain the first
integral (e.g. \citet{Spruit96}):

\begin{equation} {\rm
    Br}(\Psi)=\frac{v_{p}^{2}}{2}+\frac{(v_{\phi}-\Omega(\Psi)R)^{2}}{2}+H+\Phi-
\frac{\Omega(\Psi)^{2}R^{2}}{2}\mbox{,}\label{eq:BrFunc1}
\end{equation}

\noindent 
where $H$ is the enthalpy (defined by the equation
(\ref{eq:enthalpy}) in Appendix), $\Phi$ is the gravitational potential,
$v_p$ is the poloidal, and $v_\phi$ is the azimuthal components of the velocity.
The Bernoulli-like function, ${\rm Br[erg\cdot g^{-1}]}$ is
constant on a given flux surface $\Psi$. In the situation considered
here the gas ``lives'' on a given closed magnetic field line, in
the $\{R,\,z\}$ plane (see Figure \ref{TorusSketchFluxSurf}).  Notice that 
$\Omega(\Psi)$ is the rotation rate
of the magnetic field foot point. In a stationary toroid a closed
magnetic field line must co-rotate with the gas, i.e.
$v_{\phi}=\Omega R$ and hence with the foot point of the
corresponding magnetic field line. In general, the gas can have a
nonzero poloidal velocity, $v_{{\rm p}}$  in which case such large-scale
poloidal motions resemble meridional circulation in
stationary rotating stars \citep{Tassoul78}. Here we
assume that there is no meridional circulation. Equation
(\ref{eq:BrFunc1}) is reduced to

\begin{equation} {\rm
    Br}(\Psi)=H+\Phi-\frac{\Omega^{2}R^{2}}{2}\mbox{,}\label{eq:BrFunc2}
\end{equation}
where as before ${\rm Br(\Psi)}$ is constant
on a given flux surface.  In the original Soloviev solution
$P'(\Psi)=Const$. Making use of this insight we can assume that, since
$\Omega(\Psi)$ and ${\rm Br}(\Psi)$ are both constants on a given surface of
$\Psi=Const.$, we can make the following approximation:

\begin{figure}
  \includegraphics{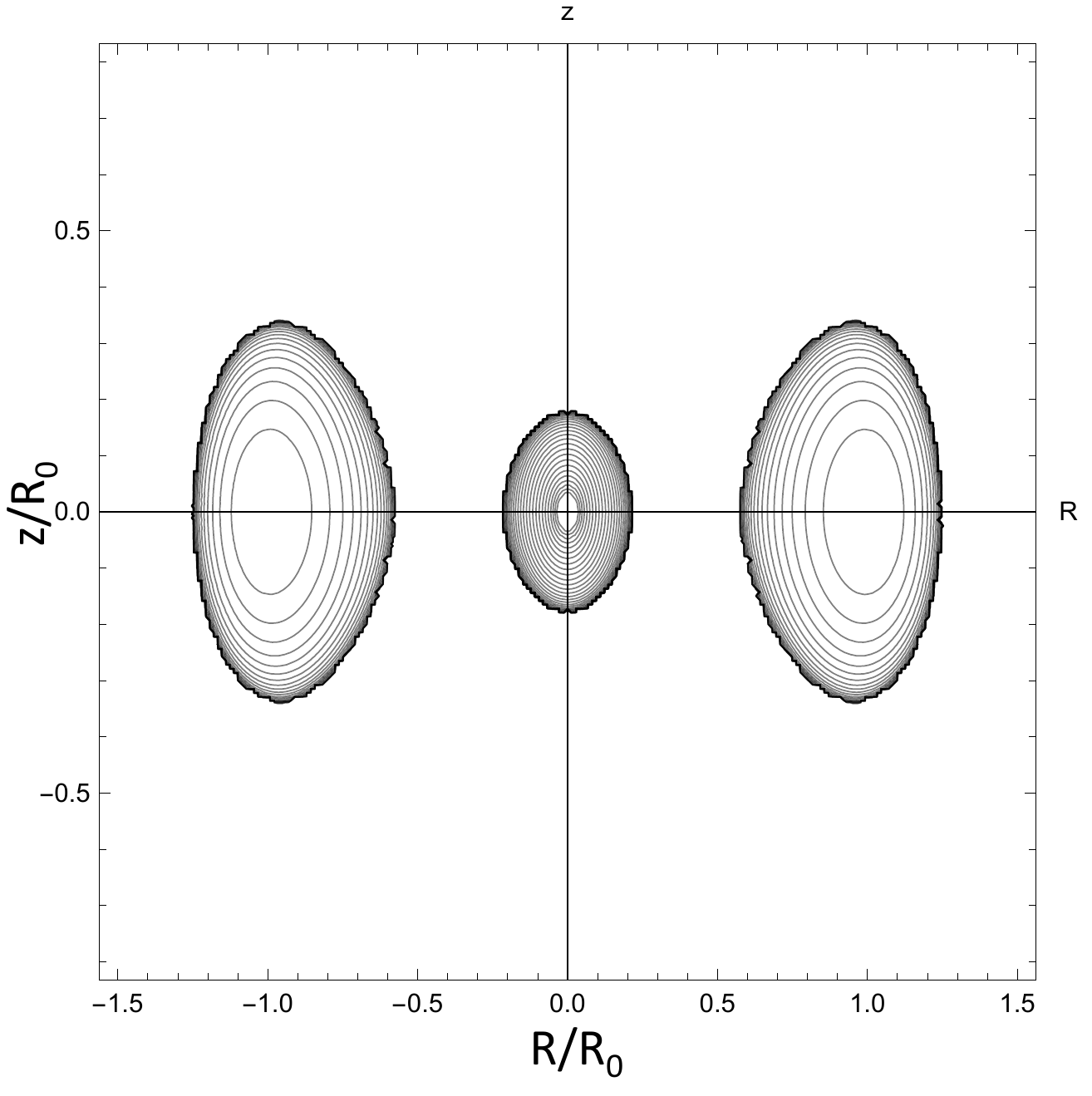} \caption{Density contours, $\log{\rho}$ 
  from equation (\ref{eq:DensitySolutionSol}) for parameters
	${\rm Br_{1}}=6$, $w_{1}=2$.
  }
  \label{Figure:AnlytSol_1Pan}
\end{figure}

For convenience we scale all the variables to their corresponding values at the equatorial plane at $R=R_0$:
$\tilde{x}=R/R_{0}$,  $\tilde{z}=z/R_0$,  and $\tilde{\rho}=\rho/\rho_{0}$; $\tilde{P}=P/P_{0}$, where
$P_{0}=\rho_{0}\,U_{0}^{2}$, $U_{0}^{2}=GM/R_{0}$, 
${\tilde \Omega}=\Omega/\Omega_0$, where $Omega_0 = U_0/R_0$;
${\rm \tilde{Br}}={\rm Br}/{U_0^2}$, $\Psi=\Psi/\Psi_{0}$  where 
$\Psi_{0}$ is an arbitrary scale factor.
We henceforth adopt these non-dimensional variables and for convenience dropping the tilde:
\begin{eqnarray}
  \Omega^{2} & = & 1 -\omega_{1}^{2}\Psi\mbox{,}\label{OmegaExpansion1}\\
  {\rm Br} & = & 1 - {\rm Br_{1}}\Psi \mbox{,} \label{BrExpansion1}
\end{eqnarray}
where $\omega_{1}$, ${\rm Br_{1}}$ are parameters.
Adopting polytropic gas, and transforming $H$ accordingly (see Appendix) we obtain from (\ref{eq:BrFunc1}):
\begin{equation}
  \rho=\frac{1}{\tilde{K}\,(1+n)}\left(1
+\frac{1}{(x^{2}+z^{2})^{1/2}}+\frac{x^{2}}{2}(1-w_{1}\Psi(x,z))-{\rm 
Br_{1}}\Psi(x,z)\right)\mbox{.}\label{eq:DensitySolutionSol}
\end{equation}
where  $\tilde{K} 
= ({\rm Br_{0}} +3/2)/(n+1)$, and $n=1/(\gamma-1)$ is the index of the 
polytrope.
A contour plot of the density in the ${x,z}$ plane obtained from (\ref{eq:DensitySolutionSol})
is shown in Figure \ref{Figure:AnlytSol_1Pan}.  It features both a
magnetized torus and a quasi-spherical region near the center. We
do not consider the latter part of the solution in the current
work.

Analytical solution of the Grad-Shafranov equation exists only in
a small number of situations which are relevant to astrophysics. When
they exist they can sometimes provide useful insight into the properties of
axially-symmetric, equilibrium configurations. Numerical solution of the 
equation can be equally complex, as the positions of the critical
points/surfaces are not known and should be found during the
solution (see the discussion in \citet{Beskin09} and references
therein).
In the subsequent sections we adopt the solution \eqref{eq:DensitySolutionSol} 
as an initial input to a full time-dependent numerical solution to the MHD equations in
an AGN torus.

\subsection{Effects of X-ray Illumination}\label{Xrays}
In luminous AGN, tori are subject to strong X-ray illumination.
Our numerical torus models include the effects of illumination, and in what
follows we discuss its role in affecting the torus structure and evolution.
To model the effects of X-ray illumination, we treat the external
continuum as being produced by the corona at the inner parts of
the accretion disk.  To isolate the effect that the X-rays play on
the dynamics of the flow, we do not consider the influence of
UV flux (see Discussion).  
Optical depth along the spherical radius $r$ is adopted to
calculate the attenuation of the incident X-ray flux:
${\tau_{\rm x} = \int \, \kappa_{\rm x}\rho \,dr }$, where
$\kappa_{\rm x}$ is the X-ray opacity which we take to be equal to the Thomson 
opacity.

Then $F_{\rm X} = F_{\rm X,0} \exp(-\tau_{\rm X})$, where $F_{\rm X,0} = L_{\rm X}/4\pi r^2$ is the local X-ray flux,
and $L_{\rm X}$ is the luminosity of the nucleus in X-rays.
We assume that a fraction
$ L_{\rm X} =f_{\rm x} L_{\rm a}$ is radiated in X-rays from the
total accretion luminosity, $L_{\rm a}=\Gamma L_{\rm edd}$, where
the latter also serves as a definition of $\Gamma$ --the important
free parameter of the problem; 
$L_{\rm edd}=1.25\cdot 10^{45}\,M_7$ is the Eddington critical
luminosity where $M_{7}$ is the mass of a BH in
$10^{7} M_{\odot}$.

 We assume that the
ionization state of the gas and the heating and cooling rates are in time-steady 
equilibrium and therefore the quantities in
(\ref{eq:energyConservEq}) are functions of the ionization
parameter:
\begin{equation}\label{smallxi}
  \xi=4\,\pi\,F_{\rm x}/n \simeq 4\cdot 10^3 \cdot f_{\rm x}\,\Gamma\, M_7/ 
(N_{23}\,r_{\rm pc})
  \mbox{,}
\end{equation}
where $N_{23}$ is the column density in $10^{23}$ ${\rm cm}^{-2}$.
The approach taken in this work with respect to X-ray heating and
cooling is similar to our previous studies
\citep[e.g.][]{Dorodnitsyn16}.

The heating and cooling function that appears in
(\ref{eq:energyConservEq}) can be written:
\begin{equation}\label{eq:HeatingAndCooling} {\cal L} =
  \frac{n}{\mu\,m_p}(\Lambda-H_{\rm X})\mbox{.}
\end{equation}
Heating and cooling rates are
calculated making use of the XSTAR photo-ionization code
\citep{KallmanBautista01} 
assuming ionizing continuum with (energy) power law index of $\alpha$=1. 
We further use analytical approximations of the above rates similar to those 
used by \cite{Blondin94}

\begin{equation}
  H_{\rm X}=H_{\rm PI}+H_{\rm IC}\mbox{,}
\end{equation}
where $H_{\rm PI}$ is the photo-ionization heating-recombination
cooling:

\begin{equation}\label{heating2}
  H_{\rm PI}({\rm erg\,cm^{3}\, s^{-1}})= 1.5\cdot 10^{-21}\,\xi^{1/4}\,T^{-
1/2}(T_{\rm x}-T)T_{\rm x}^{-1}\mbox{,}
\end{equation} 
as well as Compton heating and cooling, $H_{\rm IC}$:

\begin{equation}\label{heating1}
  H_{\rm IC}({\rm erg\,cm^{3}\, s^{-1}})= 8.9\cdot 10^{-36}\,\xi\,(T_{\rm x}-
4T)\mbox{,}
\end{equation}

The cooling rate also includes bremsstrahlung and line cooling:

\begin{eqnarray}\label{heating3}
  \Lambda({\rm erg\,cm^{3}\, s^{-1}})&=&3.3\cdot 10^{-27} T^{1/2}\nonumber\\
                                     &+&(4.6\cdot 10^{-17} \exp(-1.3\cdot 
10^5/T)\xi^{(-0.8-0.98\alpha)} T^{-1/2}+ 10^{-24})\, 
                                         \delta\mbox{.}
\end{eqnarray}
The parameter $\delta$ regulates the relative importance of
optically line cooling \citep{Blondin94}. We take the
parameter $\delta\simeq 1$ representing optically thin line
cooling.  Formulas (\ref{heating1})-(\ref{heating3}) are the
versions used by \citet{Dorodnitsyn08b}.  When compared with the
values in \citet{Blondin94} the updated results
reflect newer atomic data and the choice of a power law ionizing spectral
energy distribution.

\section{Numerical Solution\label{sec:Numerical-Solution}}

\subsection{Methods}

We further investigate the evolution of a magnetized torus exposed
to external X-ray illumination via three-dimensional numerical
simulations.  To solve a system of equations of ideal
magnetohydrodynamics
(\ref{eq:contConservEq})--(\ref{eq:magnConservEq}), we adopt the
second-order Godunov code {\tt Athena} \citep{Stone08}.  The code
is configured to adopt a uniform cylindrical grid,
$\{R_{i},\thinspace\phi_{j},\thinspace
z_{k}\}$,\citep{SkinnerOstriker2010Athena}.

Simulations are performed in a static Newtonian gravitational
potential.  The van Leer integrator is found to be the most robust
choice for the  simulations presented in this work.  We
implement the heating and cooling term in the same way as the
static gravitational potential source term is implemented in the
current van Leer integrator implementation in {\tt Athena}. That is, the heating source term
is implemented to guarantee 2-nd order accuracy in both space and time.

Given the large parameter space and numerical cost of simulations
involving X-ray heating we postpone an extensive investigation
of the parameter space to future work.  Correspondingly, we
present a comparative analysis of two models: a torus that
initially contains all of the magnetic flux inside its body; and a
torus which along with field in its body is embedded in a
large scale magnetic field of the type described by the approximate
analytic solutions to the Grad-Shafranov equation
from Section \ref{sec:ApproxSolOfTorus}.

\subsection{Initial conditions: TOR and SOL}

The two initial setups for numerical simulations 
differ according to the assumptions about the initial magnetic field
permeating the computational domain.  In the first case, the
magnetic field is completely embedded into the initial toroidal
gas configuration (hereafter TOR); In the second setup (hereafter
SOL), the initial magnetic field is calculated from the Soloviev
parametrization of the flux function, $\Psi$. 
inside the torus.
The initial
magnetization, in both cases is assumed to be small. Notice that
this assumption does not preclude the formation of highly
magnetized regions in the course of the simulation.

The TOR setup is similar to previous accretion disk simulations, 
such as \citet{Hawley2000,HawleyKrolik2001}.  Such a torus is entirely
gas pressure-supported, with angular velocity, $\Omega$ that is
constant on cylinders:

 \begin{equation}
   \Omega\sim R^{-q}\mbox{,}\label{eq:OmegaProfile}
 \end{equation}

\noindent  where, for example, $q=3/2$ corresponds to Keplerian motion
 in the equatorial plane; a torus with constant angular momentum
 corresponds to $q=2$. In cylindrical coordinates, in an axially
 symmetrical case, an equilibrium distribution of  polytropic
 gas inside rotating torus can be found analytically
 \citep{PapaloizouPringle84}:
 
%
%

\begin{equation}
  \frac{\gamma}{\gamma-1}\frac{P}{\rho}-\frac{1}{(x^{2}+z^{2})^{1/2}}-
\frac{x^{-2q+2}}{2-2q}
  =Const \mbox{,}\label{eq:torusSetupWithMagneticField}
\end{equation}

\noindent where non-dimensional variables from Section 
\ref{sec:ApproxSolOfTorus} are adopted.

\begin{figure}
  \includegraphics[scale=0.5]{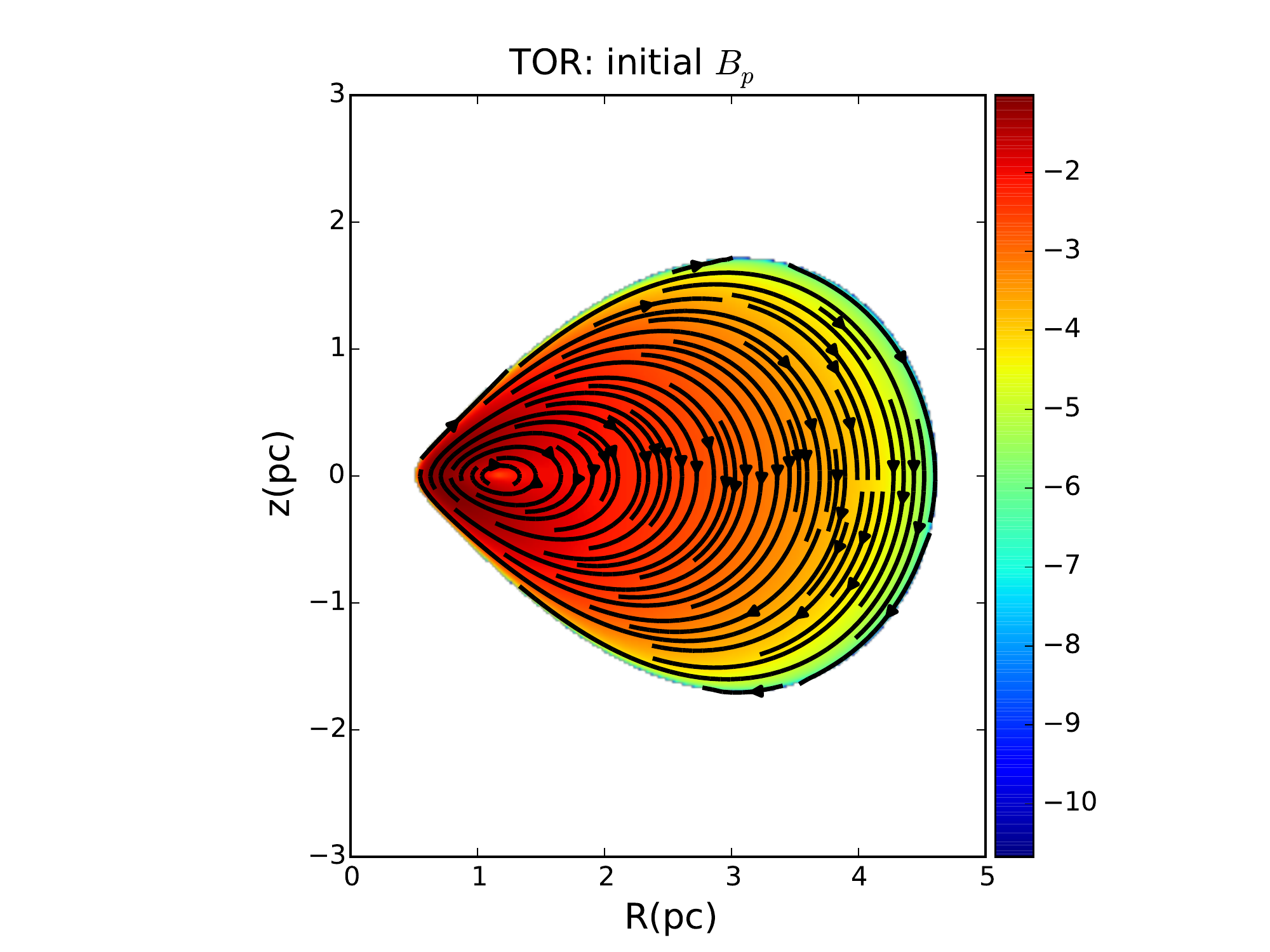}
  \caption{Initial magnetic field in the TOR simulation. Color plot: $B_{\rm p}^2/8\pi$; 
   streamlines:${\bf B}_{\rm p}$
    Axes: z: distance from equatorial plane in
    parsecs; R: distance from the BH in parsecs.}
  \label{BfieldTOR}
\end{figure}

In both the initial TOR and SOL setups only the poloidal components of the
magnetic field (i.e. $B_x$ and $B_z$) are taken into account. The
initial toroidal field in both setups is set to zero.  To
calculate $B_x$, and $B_z$ in the TOR setup, it is assumed that
inside the torus, the vector potential $A_{i}$ scales along with
the density.
In Figure \ref{Figure:AnlytSol_1Pan} and Figure \ref{BfieldTOR} the initial configurations for the B field are shown for respectively the SOL and the TOR cases.
 The use of the vector potential guarantees that
$\nabla\cdot{\bf B}=0$.  Thus in the TOR setup, the magnetic field
is calculated from ${\bf B}=\nabla\times {\bf A}$.  In the SOL
setup the magnetic field is calculated from
(\ref{eq:psiSolShort}),  i.e. from the prescribed form of
$\Psi(R,z)$ adopting the corresponding set of parameters from
Table \ref{TableParam}. Then equations (\ref{eq:bRFromPsi}) and
(\ref{eq:bZFromPsi}) are used to reconstruct $B_{r}$ and $B_{z}$.
Since $\beta_\text{m}$ depends on $\rho(x,z)$, the magnitude of the magnetic field is rescaled to
match a given value of magnetization, $\beta_{\rm m} = P/P_\text{m}$,
where $P_\text{m}$ is the magnetic pressure.  
In the TOR case $P_{\rm m}$ is initially zero outside 
the torus, so the rescaling is done only in the inside;  in the SOL case $P_{\rm m}$ is rescaled outside the torus as well as inside of it. Both simulations have the same initial $\beta_\text{m}$


\begin{table}
  \begin{tabular}{c c  c c | c c c | c c c c c c c c}
    Model &  $M_{\rm BH}/M_\odot $  &  $\Gamma$  & $f_{\rm x}$ &  $R_0$  & 
$R_{\rm in}$ & $\tau_\bot$ & 
                                                                                                                  
$\beta_{\rm m}$  & $a_1$ & $a_2$ & $b_1$ & ${\rm Br}_0$ & ${\rm Br}_1$ & $w_0$ 
& $w_1$ \\
    \hline
    \hline
    \tt  $\rm TOR $ \dotfill & $10^7$ & 0.5 & 0.5 & 1 & 0.1 & $10$ & 100 & -- & 
-- & -- & 
                                                                                            
-- & -- & --& -- \\
    \tt  $\rm SOL $ \dotfill &$10^7$ & 0.5 & 0.5 & 1 & 0.1 & $5$ & 100 & 1 & 1 & 
3 & 
                                                                                      
1& 6 & 2 & 0 \\
    \hline
  \end{tabular}
  \caption{Parameters adopted for the initial setup.}
  \label{TableParam}
\end{table}

\subsection{Input parameters}

The computational domain adopted in all the
simulations spans from $R_{{\rm in}}=0.1$ pc to
$R_{\rm out}=5$ pc and includes a $256\times 32\times 256$ in $(r, \phi, z)$ 
cylindrical grid.  
The range of $\phi$ is 0 to $\pi$ and periodic boundary conditions are imposed.
Time from our simulations is often reported in the units of the orbital
time at $R/R_0=1,$ where $R_0=1\,{\rm pc}$; that is
$t_0=1.6\times10^{3}\,{\rm yr}$.  Where appropriate we discuss the results 
referring
to the various variables cast in a non-dimensional form.
Adopted parameters are
summarized in Table \ref{TableParam} where they are shown divided
into two groups: related to the global physical properties of the
system, such as BH mass, $M_{{\rm BH}}$ and effective luminosity
parameter, $\Gamma$ and related to the characteristic physical
properties of the magnetized gas, such as density and magnetic
field.  Starting from the initially unperturbed magnetized torus,
we performed numerical simulations for several tens to several
hundreds $t_0$.
In case when the radiation pressure on dust is included \citep{Dorodnitsyn16}, the inner edge of the torus can be plausibly associated with the dust sublimation radius (see Introduction). Models discussed in the current paper do not include dust, and thus the parameter $R_0$ is a free parameter of the model.
In both TOR and SOL setups, characteristic density scale, $n_0$ 
is adjusted in the initial setup so that initial tori have similar masses, 
$M_\text{t}\sim 10^6 M_\odot$ and initial optical depth in the equatorial plane, 
$\tau_\bot$.  In the models presented here, we consider only a single value of
the parameter describing X-ray heating, $\Gamma$ - the Eddington ratio.

\section{Results\label{sec:Results}}

\subsection{General Properties}

\begin{figure}
  \includegraphics[scale=0.7]{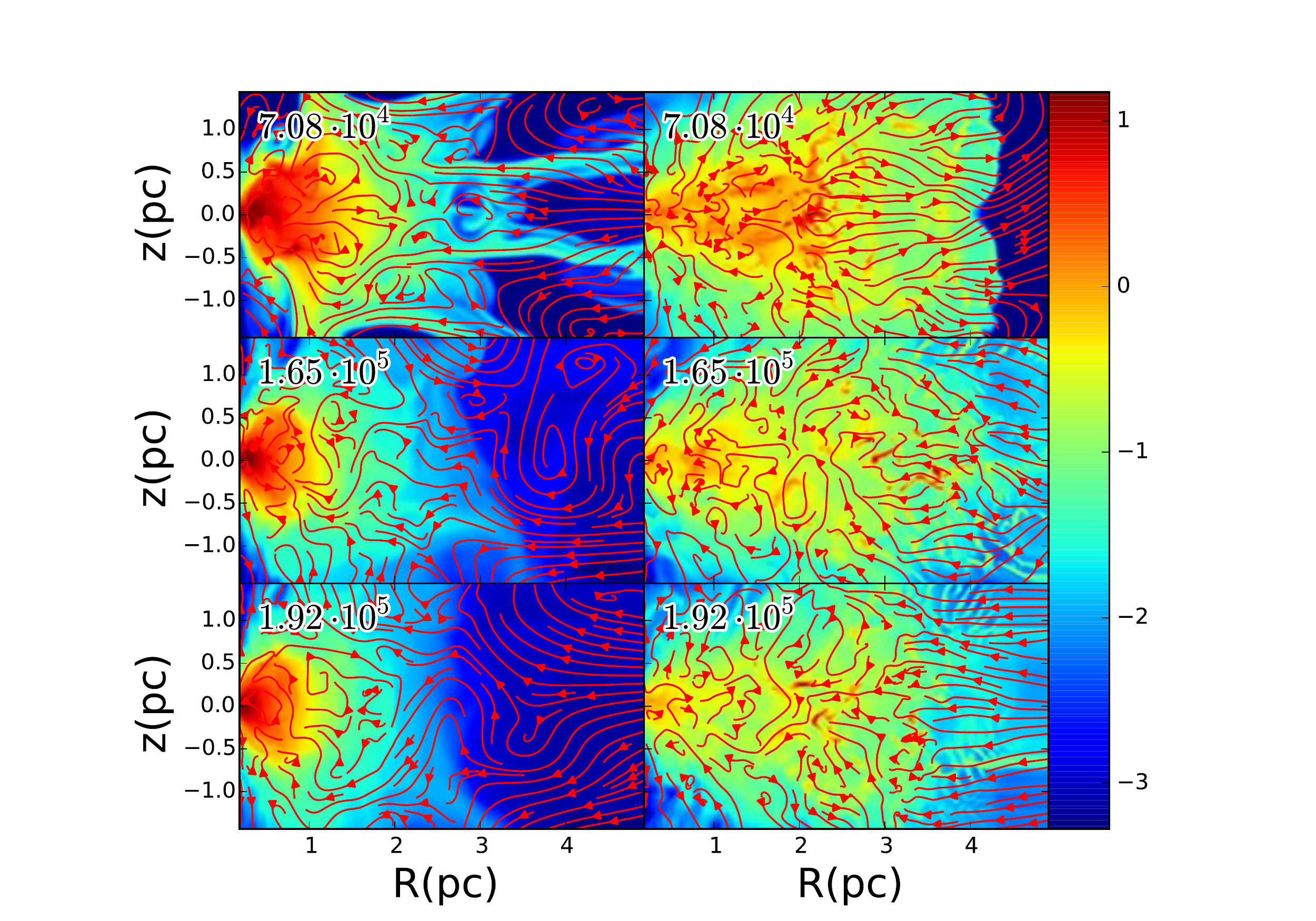}
  \caption{Color plot of the density, $\log{\rho}$ for
    $L_{\rm x} = 0.25\,L_{\rm edd}$,
    ($\Gamma = 0.5, \, f_{\rm x}=0.5$), superimposed with the velocity
    stream lines. Shown are different times given in years.  Left
    column: SOL models; right column: TOR models. Axes: horizontal: z:
    distance from equatorial plane in parsecs; R: distance from the BH
    in parsecs.}
  \label{roHWvsSOL_G05}
\end{figure}

The time evolution of our models are shown in a series of plots in Figures
\ref{roHWvsSOL_G05}-\ref{angMomHWvsSOL} where snapshots of SOL(left) runs are shown 
against snapshots of TOR (right) runs at times: $t = 15,\, 35, \text{and} \,\, 
60$. 
The different distributions of large-scale magnetic flux
inside and outside the initial torus set the SOL and TOR models on
quite different evolution tracks.  
The different evolution of the two models can be understood in terms of the relative importance
of the forces which tend to disperse the gas compared with the forces which confine it.
Radiative heating by X-rays produces temperatures which asymptotically approach the
Compton temperature, $\sim 10^7$K.  This is much greater than the virial temperature at
the torus radius, and thus leads to dispersal of the gas in the form of a wind or
super-Keplerian outflow.

Magnetic forces can disperse the torus gas due to buoyancy from field which is generated
by the MRI instability.   The strength of this process depends on the gradient of such forces
and hence on the global distribution of fields.  A key difference between the two models is
that the SOL model evolves and maintains a field strength which is approximately in equipartition
with the gas pressure over most of the computational domain.  This is true even in 
regions where the gas density is low, i.e. close to the disk rotational axis and also at large
distances from the disk plane.
The funnel in the SOL 
solution is due to the pressure of the large scale  poloidal magnetic field 
$B_\text{p}$ frozen into the gas which permeates the funnel. As the torus 
evolves, conservation of magnetic flux preserves the shape of the SOL funnel. 
The SOL model therefore preserves the shape of the gas distribution over much of the time
spanned by our simulations.  The flow 
is highly structured: the funnel clearly separates the rest of the torus from 
smaller $R$. The well defined shape of the SOL torus and the
increase of density towards the center of the torus creates a shielding layer 
which preserves the interior from strong X-ray heating.

The TOR model does not evolve equipartition fields outside the
initial gas torus region.
As the gas in the torus is also rotating
differentially the redistribution of angular momentum quickly sets
in.  
The redistribution and loss of angular momentum due to the MRI  instability and
mass loss through winds both act to change the distribution of 
gas in  the torus.  In the TOR solution the shape of the torus 
after $t \sim 10\, t_0$ bears little resemblance to its initial state.
Figure \ref{roHWvsSOL_G05} shows that in comparison with SOL models the TOR 
model has much more clumpy and filamentary structure. The funnel formed by the inner
edge of the torus is very weak or absent in case of TOR, 
Both SOL and TOR solutions have low density regions, at smaller radii. 
In the absence of viscosity or/and magnetic field the rotating gas conserves its 
specific angular momentum.
The relevance of this effect is much 
reduced in the magnetized accretion disk in which the angular momentum, $l$ is redistributed 
via magnetic stresses.

The pileup of gas at the magnetic funnel shares the same physical mechanism 
with the ``magnetically arrested disk'' \citep{Igumenshchev08}. In the following 
we will show that this difference, as expected, reduces the mass accretion rate, 
${\dot M}_{\rm a}$ for SOL solutions. Not only does the SOL solution  have piled 
up gas with higher density, larger optical depth and more complete X-ray 
screening, but also at any given time the TOR solution has lost more mass and 
thus has lower average density.  
In models which include radiation pressure similar behavior 
is observed due to the action of the radiation alone, which acts to create a well-defined
funnel by driving gas above an effective photosphere.

\subsection{Magnetic Properties}

\begin{figure}[]
  \includegraphics[scale=0.7]{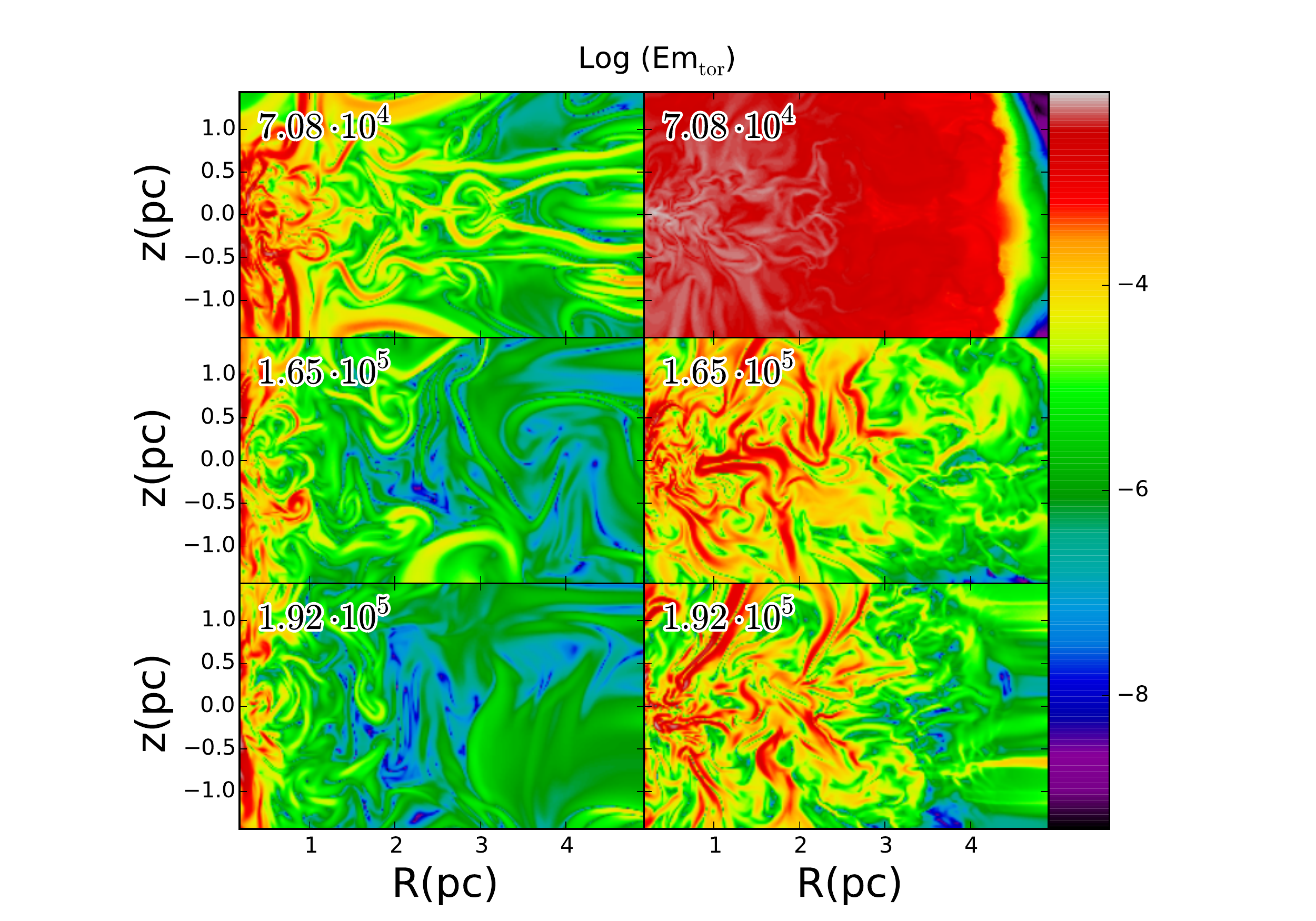}
  \caption{Energy from toroidal magnetic field
    $E_{\rm m,t}=(8\pi)^{-1}\,B_\phi^2$;
    other notation is the same as in Figure \ref{roHWvsSOL_G05}.
}
\label{magTorHWvsSOL}
\end{figure}

\begin{figure}[]
  \includegraphics[scale=0.7]{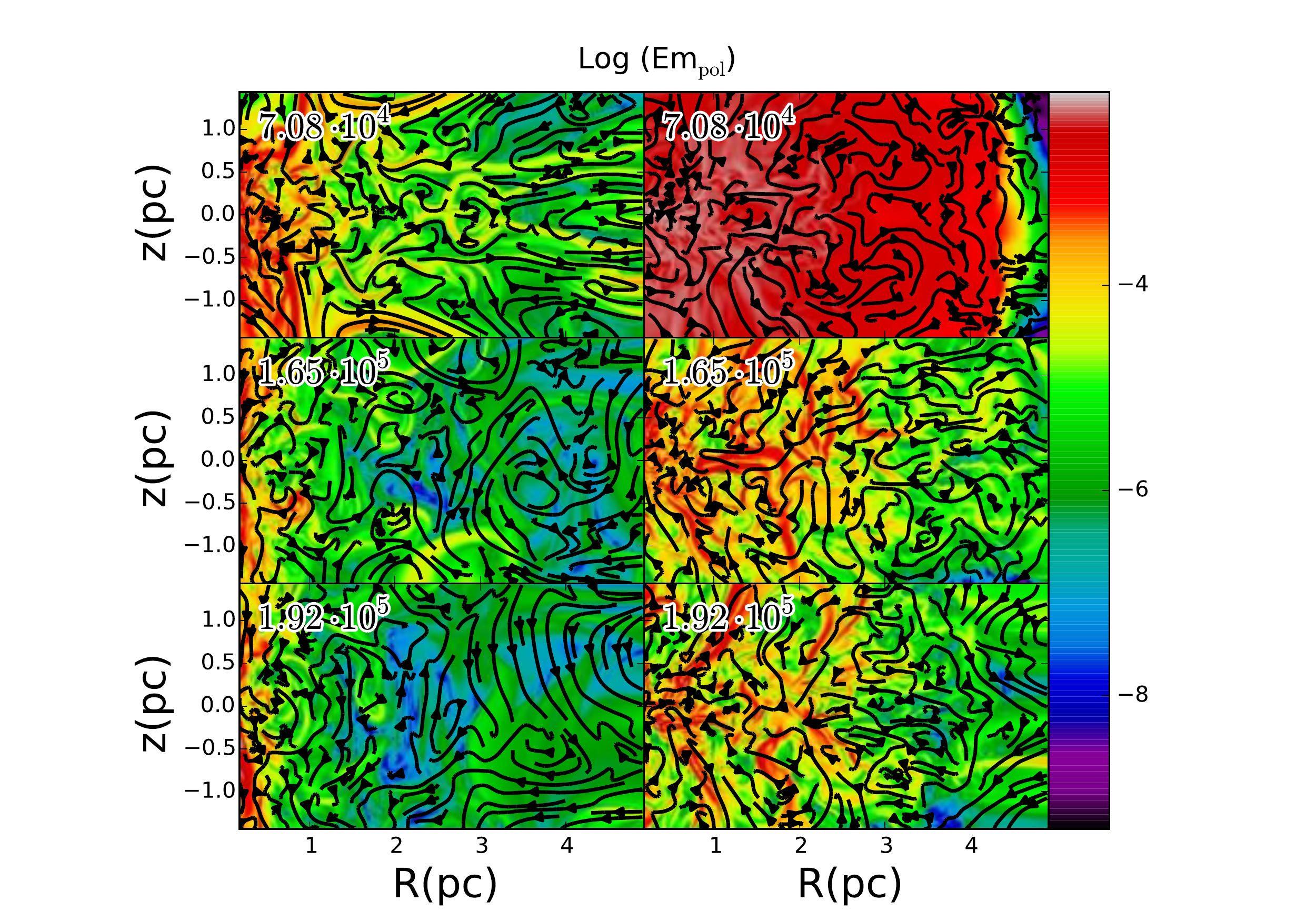}
  \caption{Energy from poloidal magnetic field $E_{\rm m,p}=(8\pi)^{-1}\,B_p^2$ with streamlines of  $B_{\rm p}$;  other notation is the same as in Figure \ref{roHWvsSOL_G05}.
}
\label{magPolHWvsSOL}
\end{figure}

In ideal MHD plasma moves easily along the
magnetic field lines while in a direction perpendicular to the magnetic field it can only move while dragging the field.  The
large-scale poloidal field is dragged towards the center until,
approximately $\rho v_r^2 \sim B_{\rm p}^2/(8\pi)$, where $B_{\rm p}$ is the poloidal magnetic field. This situation is best seen in the SOL simulation:
left panels in Figure \ref{magTorHWvsSOL} which shows the energy of the toroidal field $E_{\rm m,t}=(8\pi)^{-1}\,B_\phi^2$. Correspondingly, Figure \ref{magPolHWvsSOL} shows the energy in the poloidal magnetic field: $E_{\rm m,p}=(8\pi)^{-1}\,B_p^2$.
The total magnetic energy density,
$E_{{\rm m}}=(8\pi)^{-1}(B_{R}^{2}+B_{z}^{2}+B_{\phi}^{2})$ is
highest in and near the torus's throat and along the axis of
symmetry.

Although the TOR simulations initially contained only closed loops of $B_{\rm 
p}$, toroidal field is quickly and efficiently generated from the poloidal field
due to  differential rotation.   To estimate the production rate of 
$B_\phi$ we adopt a full form of the hydromagnetic equation \eqref{eq:rotVxB}:

\begin{equation}
  \partial_{t}{\bf B}=\nabla\times({\bf V\times{\bf B})
    -\nabla\times \left(\eta_{\rm m}\nabla\times {\bf B}\right)
    \mbox{,}}
  \label{eq:rotVxB_with_diffusivity}
\end{equation}

\noindent where $\eta_{\rm m}$ is the magnetic diffusivity - numerical or real. In the 
initial
distribution of $B$, $B_R(z=0)=0$.  Assuming the scaling $z \ll R$ near the
equatorial plane, we obtain
$ B_\phi(z\simeq 0)\sim 2\pi \frac{t}{t_{\rm sc}}\frac{R}{z} B_z $.
The field growth is eventually limited by 
diffusivity, so that $\pd_t {\bf B}\simeq 0$: From 
(\ref{eq:rotVxB_with_diffusivity})
equilibrium $B_\phi(z\simeq 0)\sim \frac{z R \Omega_{\rm K}B_z}{\eta_{\rm m}}$; 
almost everywhere else in the torus $B_\phi\sim \frac{R^2 \Omega_{\rm 
K}B_R}{\eta_{\rm m}}$,
so that toroidal field is preferentially generated from $B_r$.

In the SOL case $E_{\rm m,t}$ dominates over $E_{\rm m,p}$ in the funnel region. 
The TOR solution features a more uniform distribution over the torus region. 
An interesting feature of both solutions is the presence of low $E_{\rm m}$
filaments in the SOL solution and large-scale filaments, of enhances magnetic energy in the TOR case (Figure \ref{magTorHWvsSOL}, right).
Similar filamentary network of ``voids'' and correspondingly high $E_{\rm m,p}$ regions 
can be seen in the distribution of $E_{\rm m,p}$ in Figure \ref{magPolHWvsSOL}.
The poloidal magnetic field lines, $B_{\rm p}$ shown in Figure  \ref{magPolHWvsSOL} show a highly randomized behavior expected from the MHD-turbulent medium. Another interesting feature is that in the SOL case most of the poloidal flux is concentrated in the funnel, while in the TOR case - in the filaments.
Shearing orbital motion stretches loops of magnetic field while buoyancy and numerical magnetic reconnection produce smaller scale loops. 
Consequently, regions with the dominant $B_\phi$ are often intermittent 
with ones dominated by $B_p$.  
In pressure balance  regions where $E_{\rm m}$ is high tend to be the low density
ones. External radiation preferentially heats these low $\rho$ and high $E_{\rm 
m}$ patches. The result is a
complex pattern in the magnetic energy density distribution.
On average, $E_{\rm m,t}>E_{\rm m,p}$ in the torus. The difference of the magnetic energy
densities between the SOL and TOR simulations should be attributed
to the aforementioned stabilizing effects of large-scale
preexisting poloidal magnetic flux.  Comparing Figure \ref{roHWvsSOL_G05} to 
Figure \ref{magTorHWvsSOL} and Figure \ref{magPolHWvsSOL} one can see that 
in general, in the SOL solutions, regions of high $E_{\rm m}$ are the regions 
of increased density.  

\subsection{Thermodynamics and effects of X-ray heating}

\begin{figure}
  \includegraphics[scale=0.7]{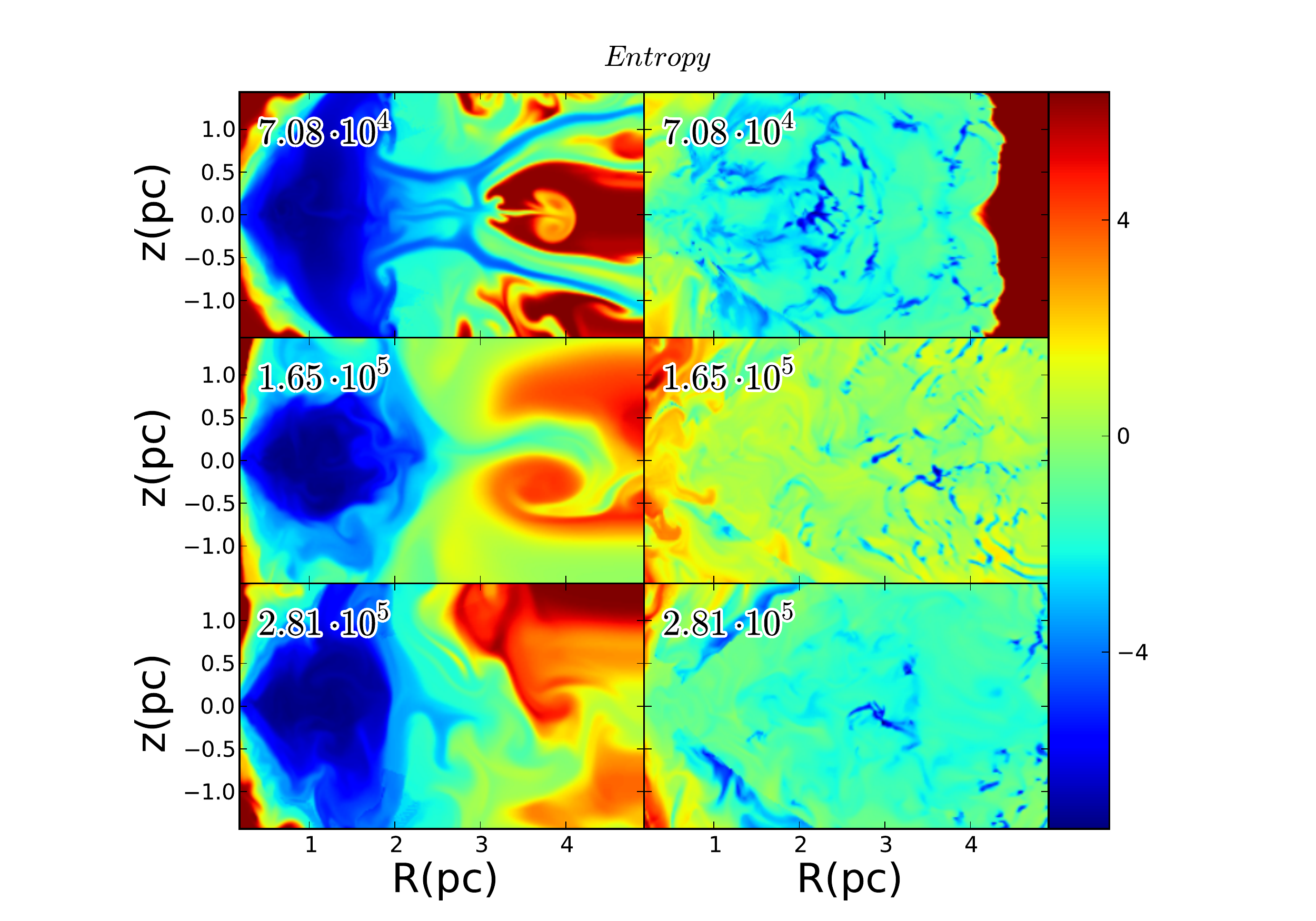}
  \caption{Specific
    entropy; other notation is the same as in Figure \ref{roHWvsSOL_G05}.}
  \label{entrHWvsSOL}
\end{figure}

Figure \ref{entrHWvsSOL} shows the distribution of the
non-dimensional entropy, $S=\log P/\rho^\gamma$.  Low entropy gas
fills the torus in the SOL case suggesting a smaller extent of mixing 
during the evolution. Low density, high
entropy gas is found near the rotational axis and at larger radii
where large scale gas motions occur. 
The gas piled up near the magnetic funnel in the SOL solution is dense, and 
correspondingly the entropy is low. At the same time the TOR solution has lost 
more mass and the radiation heating is more uniform across the range of radii.
In the TOR case the entropy
distribution shows filamentary structure.  
Comparing Figure \ref{roHWvsSOL_G05} to Figure 
\ref{magTorHWvsSOL} and Figure \ref{magPolHWvsSOL} one can see that
SOL torus has markedly lower entropy than that of the TOR. In the TOR case,
entropy traces the filamentary and clumpy structure of the distribution of the
average magnetic field.

Different topologies of the initial field results in different levels of MRI-
driven  stirring of the torus interior.
The density near the disk funnel is naturally higher than the 
density 
in a smooth disk of the same mass. Shielding depends on the optical depth of the 
obscuring
region, $\sim \bar{\rho}\, \delta l$. In the SOL solution there is little gas in 
the central hole, and thus the characteristic length scale, $\delta l$ is 
smaller, but this is more than compensated  by the increase of the density in 
this region, $\bar{\rho}$ due to  compression of the funnel walls. 
Correspondingly, the SOL solution shows  more  efficient shielding from X-rays. 
Thus,
the gas in TOR solution is considerably hotter, $T\simeq 10^5$K than in SOL 
case, 
$T \simeq 10^4$K. Near the axis the gas is very hot, almost virialized: in SOL 
case
it is $T \simeq 10\times 10^7$K and in TOR case it is $T\simeq 2.5\times 
10^6$K. 
Angular dependence of the optical depth manifests itself in ``ionization cones''
(Figure \ref{entrHWvsSOL}, right panels).

Extensive
irradiation by X-rays along with the radiative cooling enhance density 
contrasts produced from MRI-driven convection. 
Tests of non-irradiated tori show that 
the conclusions about the effects of magnetic field topology in the SOL 
vs TOR setups remain qualitatively unchanged.  However, 
X-ray heating and cooling create more turbulent and irregular tori as 
a result of enhanced convection.

\begin{figure}
  \includegraphics[scale=0.7]{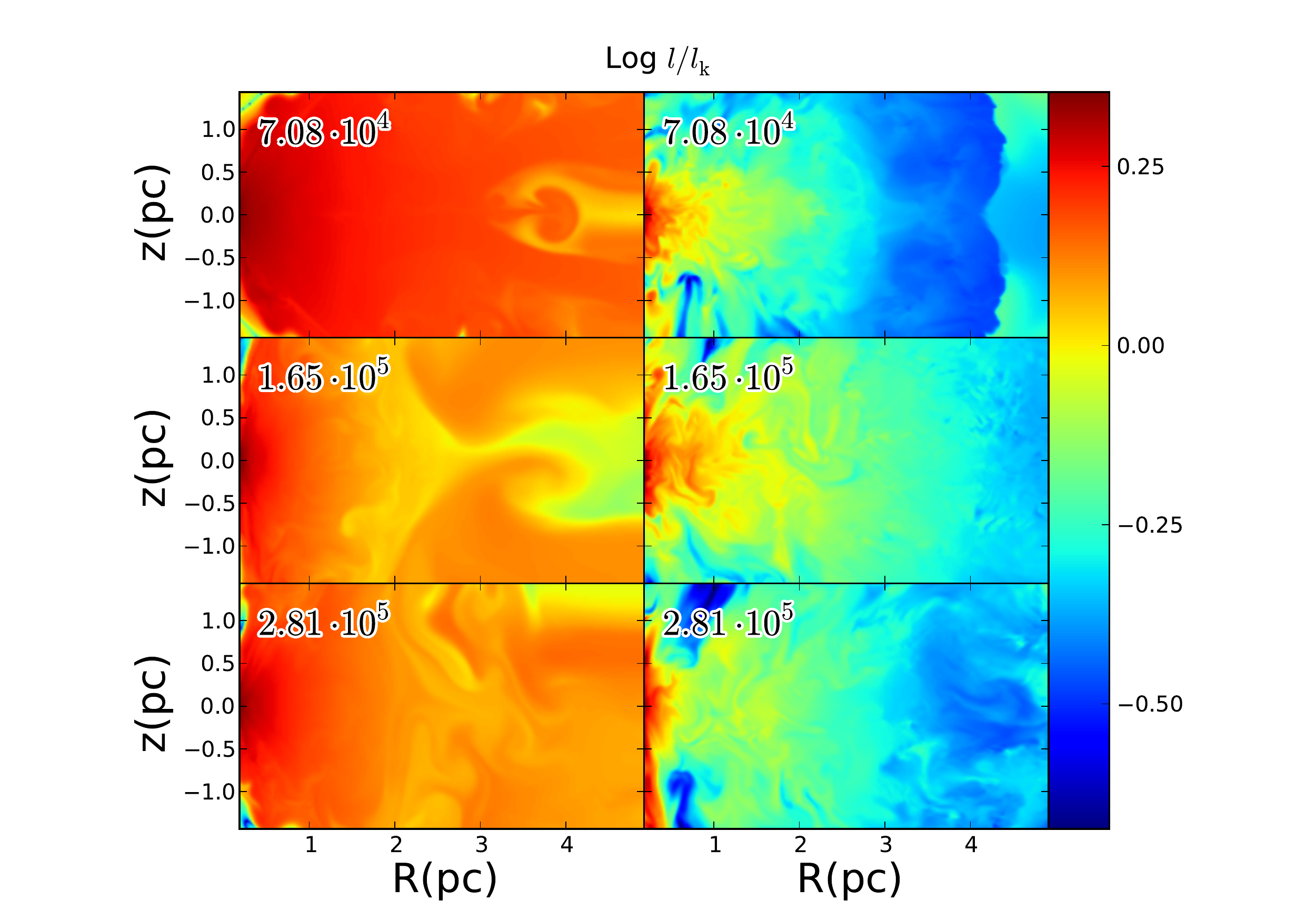}
  \caption{Angular momentum: $l/l_{\rm k}$, where 
    $l_{\rm k}$ is the Keplerian angular momentum.; other notation is the same 
as in Figure \ref{roHWvsSOL_G05}. }
\label{angMomHWvsSOL}
\end{figure}

\subsection{Velocity and Rotation}

The large scale magnetic field in the SOL simulation is effective in smoothing 
the contrast in specific angular momentum $l$, so that this quantity remains 
close to its local Keplerian value, $l_{\rm k}$.
This is not true for the TOR simulations.
Figure \ref{angMomHWvsSOL}  shows the distribution of $l(R,z)$.
In the TOR solutions, $l$ spans the full range, from $l\sim l_{\rm k}$ near the 
inner domain boundary to an approximately half of $l_{\rm k}$ at larger $R$.
 The rotation velocity, $v_{\phi}$
remains nearly Keplerian inside a significant part of the
torus. This is more true  close to
the equator. In the torus throat most of the low density gas is
sub-Keplerian. However, the gas in this region is too dilute to
have a noticeable influence on the dynamics and torus.
Figure \ref{roHWvsSOL_G05} shows an  over-dense stream protruding to
small $R$ at late times as the result of the formation of an
equatorial, nearly Keplerian, small accretion disk at late stages of the
simulation.

Streamlines of the poloidal velocity, $v_{\rm p}$ are shown in Figure 
~\ref{roHWvsSOL_G05}. In the funnel region, the SOL solution has  $v_{\rm p}$ 
change from inflow (upper left) to outflow (lower left); inside the main body of 
the torus, the large scale motion in the SOL solution resembles meridional flows 
with axial symmetry, but with no equatorial symmetry. In the TOR case the 
velocity pattern is highly chaotic, without any obvious implied symmetry,
though there is a net outflow in the regions near the lower and upper boundaries.



\subsection{Obscuration}

An important, potentially observable characteristic of a torus is
its obscuring properties.
The distribution of gas in the computational domain depends on the gas rotation 
(centrifugal barrier), radiation pressure, and magnetic pressure and tension.
When gas has means to shed its angular momentum the centrifugal barrier becomes 
less relevant.  In the case such as ours,  when there is no radiation pressure, 
the role of 
the large scale magnetic field can be disentangled from other forces shaping the 
torus. Figure \ref{figObscuration} shows column density 
$N_\text{col}(\theta)= \int_s\,n(r,\theta)\,dr$, as a function of the 
inclination angle, $\theta$ measured from the axis of rotation. Each point of a 
particular color indicates a model at time step, $t$ i.e. as recorded over the 
entire simulation time range. There are total of 500 model shown for each setup 
(SOL: left, or TOR: right).

The qualitative  behavior of the solutions is consistent with the Figure 
\ref{roHWvsSOL_G05}:
The torus confinement by the pressure of the magnetic funnel is quite pronounced 
in the SOL (left) setup where $N_\text{col}(\theta)$ practically tracks the 
torus funnel where
the gas is piling up. In the TOR setup the distribution of 
$N_\text{col}(\theta)$ is considerably more uniform.  Gas that fills the torus 
throat provides
only negligible obscuration due to its low density. These results demonstrate 
that
the magnetic torus  can provide enough obscuration, and which lasts for enough 
time, to be
relevant for the problem of Type II AGN obscuration.
supports accretion rate close to the maximum efficiency. From Figure \ref{figObscuration}
the TOR solution provides larger columns near 90 degrees inclination. 
The magnetic 
funnel in the SOL case provides a barrier where the gas is piling up. Thus, at smaller inclinations 
the SOL disk tends to thin out less at later times, than the TOR disk which does not have such a 
magnetic funnel.

\begin{figure}
  \includegraphics[scale=0.4]{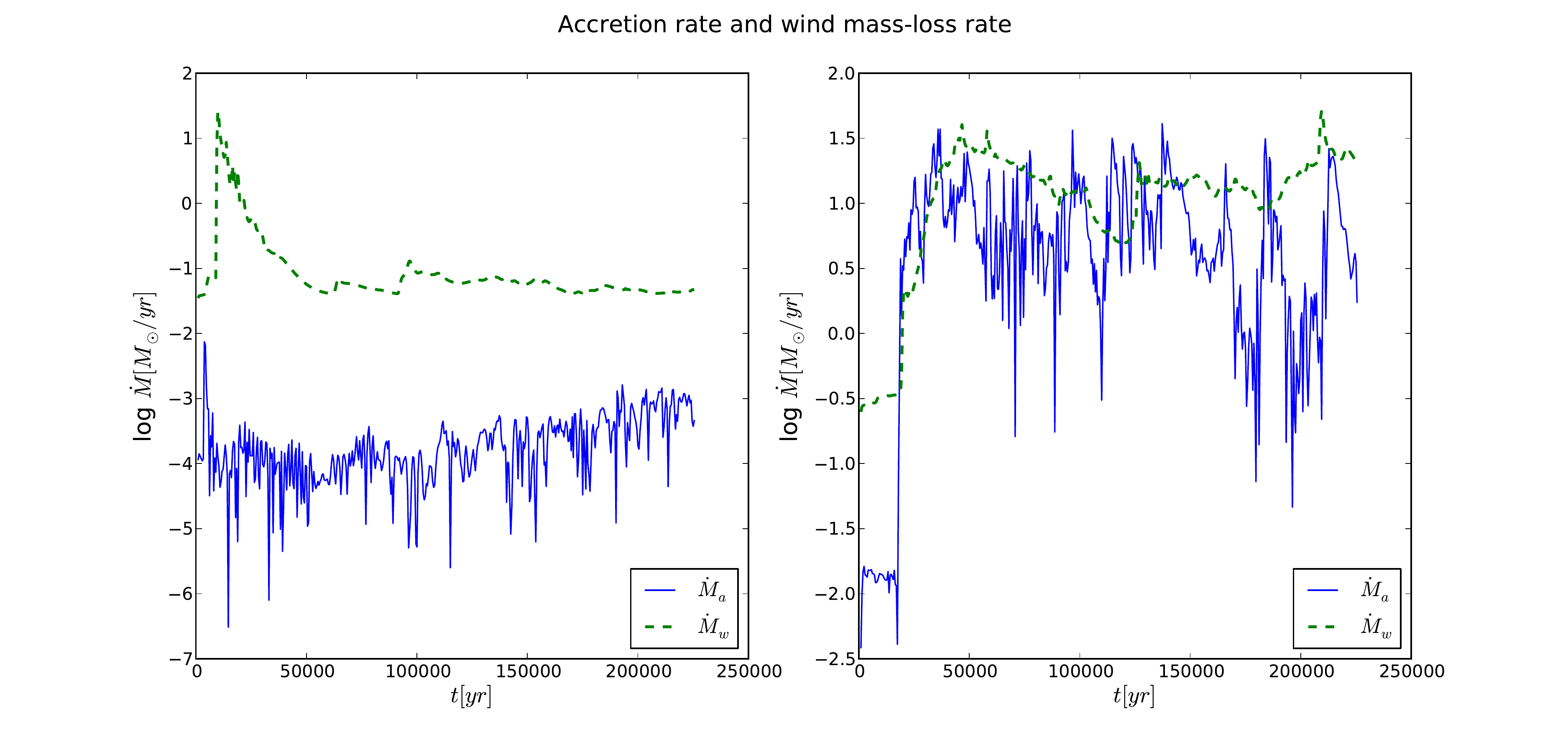}
  \caption{Accretion rate, $\dot{M}_{{\rm a}}$ and the total wind
    mass-loss rate, $\dot{M}_{{\rm w}}$ versus time for two
    different initial setups: Left: SOL; right: TOR}
  \centering{}\label{Figure MdotAccr}
\end{figure}

\begin{figure}
  \includegraphics[scale=0.4]{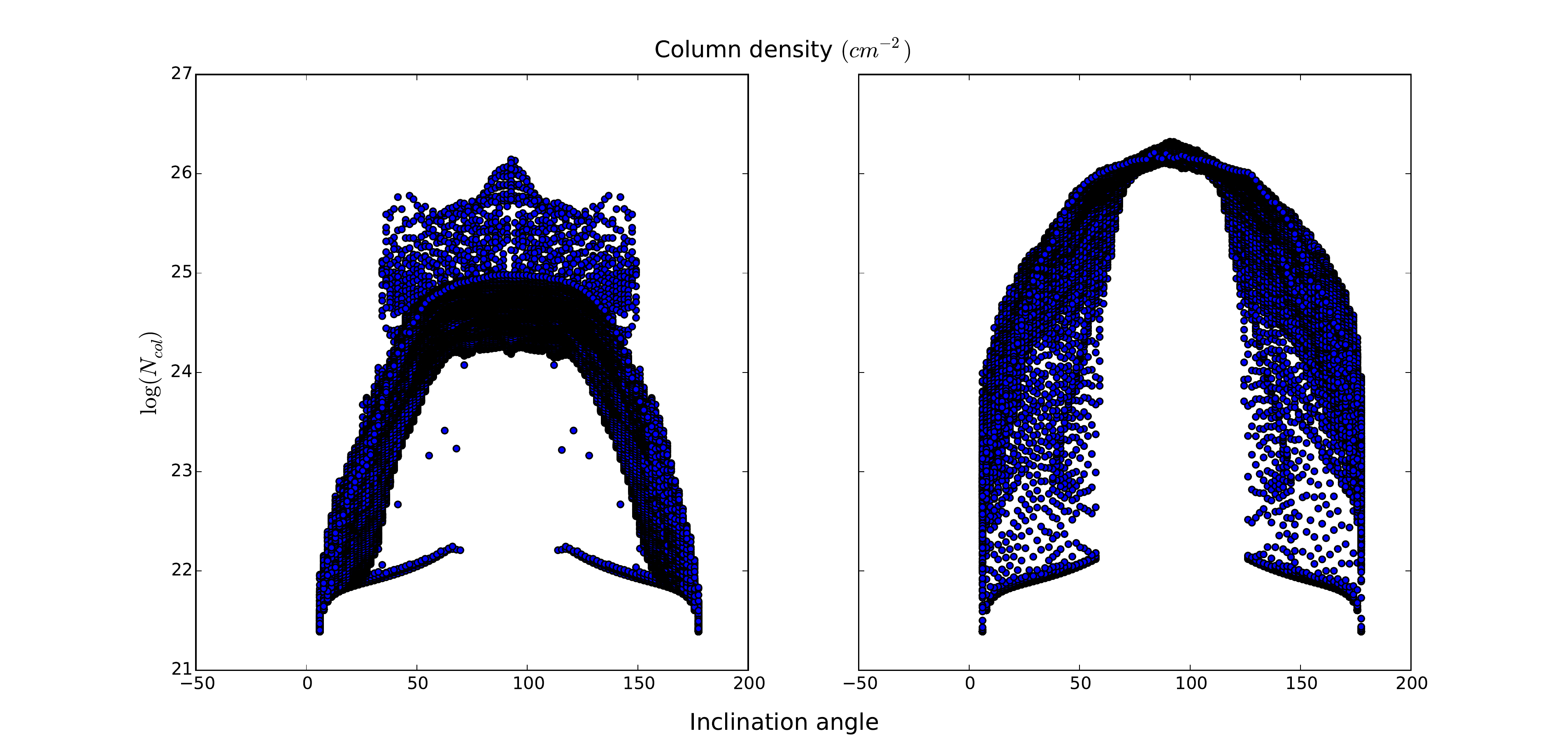}
\caption{Scatter
    plot showing the logarithm of the column density,
    $\ensuremath{N_{{\rm col}}(\text{cm}^{-2}})$.  Each recorded
    simulation time is represented by a point. Left: SOL; right: TOR}

  \label{figObscuration}
\end{figure}

\subsection{Accretion and Wind Mass Loss}

Gas can be lost from the system through excretion or
wind. Figure \ref{Figure MdotAccr} shows mass-loss rates through
accretion, $\dot{M}_{{\rm a}}$ and through the wind
$\dot{M}_{{\rm w}}$ for the SOL(left) and TOR(right) setups.
After an initial transient period a
quasi-stationary regime of accretion is established in both cases.

Wind mass loss occurs from the upper, lower and right
(furthermost from the BH) domain boundaries.  Mass loss rates from the
upper, $\dot{M}_{{\rm w}}^{+}$ and lower $\dot{M}_{{\rm w}}^{-}$ dominate the outflow rate
while the outflow at the outer $R$ direction, after initial short increase becomes negligible, less than $10^{-10}\Msy$. An apparent inflow at the right boundary in Figure \ref{roHWvsSOL_G05} occurs in the 
region of a very low density and thus carries very little mass.

Within the time span of the simulation, the torus
sheds more mass via $\dot{M}_{{\rm w}}$ than from
$\dot{M}_{{\rm a}}$. Excretion is expected as angular momentum is
transferred outwards by both the wind and equatorial outflow.
After $T\sim 2\times 10^4\, \text{yrs}$ in both runs the mass-loss rate 
enters a regime in which $\MdotAccr$ is fluctuating within a range; in case of SOL: 
$\dot{M}_{{\rm a}} \simeq 10^{-4}- 10^{-3}\,\Msy$, and 
TOR: $\dot{M}_{{\rm a}} \simeq 10^{-1}-3\times10^{1}\,\Msy$. There is no obvious increasing or decreasing trend in average mass-accretion rate,  $\langle 
\MdotAccr \rangle$. The largest variability scale of the TOR is about $10^4 \,
\text{yrs}$ while in the SOL it is much smaller, about $10^3\,\text{yrs}$.  The 
variability amplitude in the SOL solutions is
generally smaller than in the TOR solutions.
In both cases the wind mass-loss rate, $\dot{M}_{{\rm w}} \simeq 0.01-0.1\Msy$. 
Notice that for our model parameters, the Eddington mass-accretion rate, ${\dot 
M}_\text{a,Edd}  \simeq 0.1\Msy$
where we assumed the efficiency of accretion, $\epsilon=0.1$.
The accretion rate in SOL and TOR runs almost always stays below this level with 
the excess energy removed by  winds.

\subsection{Comparison to the radiation-pressure supported torus}

It is instructive to compare the obscuring properties of the
radiation-pressure supported model of the torus from
\citet{Dorodnitsyn16} with predictions of the magnetized
torus.  In the magnetic case the torus evolves more slowly.
much. In a radiation supported scenario, even at a relatively
low radiation energy input of $L=0.01\,L_{{\rm edd}},$ where
$L_{{\rm edd}}$ is the Eddington luminosity, at comparable time
since the beginning of the simulations, the torus is significantly
more distorted, and windy (see Figure 1 from
\citet{Dorodnitsyn16}). In the SOL or TOR cases an organized wind is much weaker than in the radiation pressure case.
The organized pattern of streamlines is found mostly
outside the torus, in the region of tenuous plasma.

Comparing properties of magnetic tori with the gas-only 
and radiation pressure supported tori having similar density scales, $n_0=10^8$,
shows that the torus mass, $M_\text{t}\simeq 10^6M_\odot$ is similar to the gas-only torus
\citep{Dorodnitsyn08b}: $M_\text{t}\simeq 9.3\times 10^5M_\odot$, and in 
the IR-pressure supported torus $M_\text{t}\simeq 5\times 10^5M_\odot$,
\citep{Dorodnitsyn16}. The mass-loss rates, however, are noticeably higher in 
the radiation pressure supported model. For example, at a comparable radiation input level,
$\Gamma=0.3$ the mass-loss rate in the IR-supported torus is $\dot{M}_{{\rm w}}\simeq 0.2\Msy$.

\section{Discussion\label{sec:Discussion}}


We have previously conducted a series of hydrodynamic simulations of the torus
designed to better understand its relation 
to the problem of Seyfert II obscuration and AGN unification
 as well as its relation to the inner accretion disk.
In the current work we report our first results on three-dimensional modeling of 
magnetized AGN tori exposed to external X-ray heating. 

We have found a connection of this problem to the problem of magnetically 
supported tori in Tokamaks.
In the context of Tokamak studies axially-symmetric solutions for a non-
rotating, magnetically confined and supported torus made of polytropic gas 
exists in the literature.
Our analytical estimates included the analysis of the solutions to the Grad-
Shafranov equation.  We obtain an approximate solution to this problem that 
demonstrates the possibility of torus be support by the poloidal magnetic field. 

In a realistic differentially rotating torus the effects of MRI-driven 
turbulence are expected to transform the torus into an accretion disk on a 
dynamical scale. 
The results from our numerical simulations confirm this expectation.  Evidence 
from this and other simulations \citep{Beckwith09,Hawley2000} suggests that 
there is an important difference in the outcome of the accretion depending on 
how the initial magnetic flux is configured. 

After the MRI-stirred turbulence sets in, the
magnetic field can be viewed as being composed of a regular $\langle {\bf B} 
\rangle$ and a random
 ${\bf B}'$ components:

\begin{equation}
{\bf B}= \langle{\bf B} \rangle +{\bf B}' \mbox{,}
\end{equation}

\noindent where large scale $\langle {\bf B} \rangle$ is averaged over the statistical 
ensemble that involves a fluctuating component of the magnetic field, ${\bf 
B}'$. 
If  the poloidal component in the initial  large-scale {\bf B} is strong enough 
(as in our SOL setup) previous simulations  find that typically
$\langle {\bf B}_{\rm p} \rangle \gtrsim {\bf B}' $ and poloidal flux can be 
accumulated  at smaller radii creating a magnetic bundle that impedes 
accretion. Higher mass fluxes, on the other hand, can enhance mixing between 
regular and spatially chaotic motions, leading to
$\langle {\bf B}_{\rm R} \rangle  \sim \langle {\bf B}_{\rm z} \rangle$.

It is likely that during a significant episode of accretion, such as may happen 
during 
a major merger for example, a significant bundle of large scale magnetic field 
will be dragged from galactic scales to $\sim$pc scales.
Orbital shear and amplification due to flux freezing can further shape the 
toroidal topology of such a field.
If initially most of the $B_p$-flux is contained inside a torus then the results 
of our TOR simulation shows that
it will lead to a  more spatially chaotic accretion with no magnetic funnel.
If significant flux is carried by low-density gas then the results of our SOL 
simulation shows that an inner magnetic bundle forms.

One should exercise caution in assessing the mechanism of the vertical magnetic 
field transport. Important collective insight from 
theoretical and numerical work \citep{BKLovelace2007,Beckwith09} demonstrate 
that low density, high $z$ regions of the disk are the main provider and 
contributor of the advection of $B_{\rm p}$. The equatorial part of the disk is 
turbulent and effectively acts as a diamagnetic, rendering frozen in condition 
irrelevant in this layer and instead allowing matter to slip through the field 
lines.  
The accumulation of the poloidal field in this low density regions at smaller 
radii, creates a magnetically arrested disk (MAD) state.
The equatorial flow is still reminiscent of thin, turbulent disk. It is in this 
thin layer where accretion is mostly happening in the SOL case despite the MAD 
state of the disk as a whole.

To assess whether or not the MRI-turbulence is numerically resolved, one 
compares the fastest-growing MRI mode, $\lambda_{\rm MRI}$ with the size of the 
computational grid cell. The safety factor is $Q_\phi=\lambda_{\rm MRI}/(R 
\Delta\phi)$ \citep[e.g.][]{Noble2010, Sorathia12}, where $\Delta\phi$ is the 
resolution in the $\phi$ direction. In our notation it is, then 
$Q_\phi=2\pi\sqrt{2} ( P/(\rho\beta_{\rm m} )^{1/2}/(V_{\rm k}(z=0) 
\Delta\phi)$, where $V_{\rm k}(z=0)$ is the Keplerian velocity at the equatorial 
plane. In our simulations $Q_\phi \simeq 1-5$.
Our simulations do not resolve MHD turbulence, thus all the effects associated 
with magnetic diffusivity $\eta_{\rm m}$ should be attributed to the numerical 
diffusivity of the method. 
However numerical diffusivity is working towards the same goal as it would be in 
the case of the effective turbulent diffusivity, $\eta_{\rm m}$: magnetic field 
diffuses through the accreting gas.

The primary role played by X-ray heating in our simulations is that it enhances 
entropy contrasts, since the heating is greater in low density regions.
In the case of the magnetized inner funnel this leads to enhanced gas 
evaporation. 
In previous papers we found that incident X-rays heat the surface of the torus 
creating a thin overheated layer (``a skin'') which forms a base for an 
evaporative outflow. 

Our simulations include the effects of X-ray heating on the torus gas.
The primary role played by X-ray heating  is that it enhances 
entropy contrasts, since the heating is greater in low density regions.
In the case of the magnetized inner funnel this leads to enhanced gas 
evaporation.   This paper does not include any of the various other
effects of radiation in torus models, including the effects of
X-rays and UV in dust grain photodestruction, radiation pressure on
dust and on free electrons from UV and on dust from X-rays,
dust heating by UV and X-rays and reprocessing into
infrared radiation, and the associated pressure from the infrared in a dusty medium.
In the models described in this paper all the vertical support comes from the 
gas and magnetic pressure and tension. 
The torus vertical support and confinement is the result
of the combined action of the gas pressure and the magnetic field pressure and tension.
The magnetic field in this scenario is dragged along with the gas, from the galactic scales.

Though the effects of dust and IR are not included in this paper, it is clear that the survival
of dust against evaporation by X-rays is a crucial determinant of the dynamical effects of
radiation.  For instance, if a cold slab of plasma is exposed to 
unattenuated X-ray and UV radiation, a significant part of such radiation
will be absorbed and reprocessed into infrared 
within a layer of thickness, $\delta l/R_{\rm 1pc}\simeq 1.3\times {\rm 10^{-3}} (n/10^7{\rm cm^{-3}})$.
This will create  radiation pressure, ${\bf g}_r$ which is primarily in the (spherically) outward
direction.
The critical Eddington luminosity with respect to dust opacity is
$\Gamma_{\rm IR} = \alpha\, \Gamma \eta_{\rm X} \kappa_{\rm d}/\kappa_{\rm e} \simeq 1.25$ 
for $M_{\rm BH}=10^{7}\,\text{M}_\odot$,  $L=0.5 L_{\rm edd}$ and $\kappa_{\rm d}=10 \kappa_{\rm e}$; also adopting $\eta_{\rm X}=0.5$ for the fraction of X-ray radiation from the total radiation, and the fraction $\alpha\simeq 0.5$ of the incident flux is re-emitted outwardly.
In our previous  papers \citep{Dorodnitsyn16,Dorodnitsyn12b}
we have included the effects associated with dust and X-rays:  the conversion of
X-rays to infrared, and the IR radiation pressure.  
In the UV the dust opacity is $\geq 10^2$ times the  Thomson opacity, and is correspondingly
greater  than the X-ray opacity.  This results in the absorption of UV in a much thinner 'skin'
layer than is the case for X-rays.
Far from the disk plane, and at low density, dust evaporation dramatically
reduces the vertical component of ${\bf g}_r$.
Close to the disk plane the UV attenuation layer is very thin.
The combined pressure of UV radiation and of the pressure of the 
hot gas on the torus's inside funnel wall pushes the torus outwards, increasing  
$R_0$.  Quantitative exploration
of these effects will be undertaken in a later paper.

It was also found in \citet[][]{Dorodnitsyn16} that the formation of a
funnel can significantly slow down 
or even stop  accretion via disk viscosity. However, it was found 
that in this situation
accretion can still proceed through capturing the thermally evaporated wind 
while the accretion disk remains truncated by radiation pressure and heating.

So far we have also neglected non-ideal effects: non-ideal plasma effects can 
play an important role deep inside the torus where the gas can be cold and 
neutral and
there may be not enough charge-carrying particles. 
If so, magnetic diffusivity in the equation (\ref{eq:rotVxB_with_diffusivity})
will be significant, and will allow the magnetic field to effectively slip 
through the moving
fluid. Reduced magnetic dragging means less magnetic force. 
It is well known that inside a thin accretion disk 
turbulence can drastically increase the effective diffusivity which renders the 
transport of the
large-scale magnetic fields nearly impossible
\citep{LubowPapaloizou94}.  Very low physical conductivity inside a cold torus 
will work the same way
as high effective diffusivity  in accretion disks.
On the other hand, it has been shown that radiative
layers in the disk \citet{BKLovelace2007} and corona,
and/or  low density conductive regions restore the
coupling \citep{BKLovelace2000,Beckwith09}.
High effective diffusivity inside the torus imply that dynamically
important fields should be transported by the hot conductive, diluted gas in
which the torus is in pressure equilibrium.  The torus will be
embedded into such gas and this gas  will carry most of the current and
magnetic field which in turn will play important role in confining
the torus. In our previous simulations X-ray illumination produced large
via evaporation of the cold torus.
Our assumption  that the gas is sufficiently ionized to
validate the use of ideal MHD is likely to be applicable
whenever the electron fraction exceeds $\ensuremath{\sim10^{-11}}$
\citet{Fleming03}.  The temperatures in our previous IR-driven tori
so far are $\ensuremath{\geq}1000{\rm \,K}$. This insures that there is enough 
charge carrying 
particles to justify flux-frozen condition.

\section{Conclusions\label{sec:Conclusions}}

We performed numerical simulations of AGN tori threaded  by
a large scale magnetic field.
To illustrate the feasibility of this model, we first studied an illustrative 
analytical example of the torus that is supported and confined by the strong 
poloidal, large-scale magnetic field. To isolate the effects of the
magnetic support, our three-dimensional numerical model is decisively 
simplified and 
includes only large scale  magnetic field along with the gas
pressure and  X-ray heating and cooling. Radiation pressure in any form is 
neglected.  We argue that despite numerous simplifications, the results are in 
fact in a good agreement with
observed properties of AGN tori, and that they
validate a scenario in which the torus is a vertically thick accretion disk whose
evolution is  governed by
the magnetic flux supplied from galactic scales.
The magnetic field has a major effect on the torus 
morphology and dynamics: 

\begin{itemize}
\item 
We derive an analytic model  by considering a generalization 
of the magnetic topology known in a Tokamak literature as a ``Soloviev 
solution''.
Our magnetically confined torus features a poloidal magnetic field which is 
partly trapped inside the torus where it contributes to its vertical support 
against gravity. This model also has a significant magnetic flux outside the 
torus, in the low density gas. 
We argue that both of these initial states may result from accretion of 
magnetized gas from galactic scales
in which case a dynamically significant magnetic flux can accumulate in the 
inner regions of AGN as a result of magnetic field dragging from galactic 
scales. 
\item
Two types of initial magnetic topologies are  studied: in the first one (SOL)  
the initial state is constructed from the modified Soloviev equilibrium. In the 
second setup (TOR), the magnetic flux is initially contained inside the torus in 
which case such simulation shares many common features with corresponding 
simulations of geometrically thick accretion disks. In both cases plasma is 
exposed to external X-ray illumination, heating and cooling.
\item
Two main results arrive from these simulations: i) TOR  setup evolves into a 
geometrically thick convective/turbulent accretion disk with $\langle {\bf 
B}_{\rm R} \rangle  \sim \langle {\bf B}_{\rm z} \rangle$. 
The magnetic field provides a mechanism to redistribute angular momentum 
through MRI turbulence and outflows.  
ii) The SOL case evolves into a configuration that resembles the arrested 
 disk with a strongly magnetized torus funnel. 
\item
External X-ray irradiation does not qualitatively change the results; however 
extensive heating does shift the outcome to a more turbulent and irregular torus 
as a result of enhanced convection.
\end{itemize}

\acknowledgements  
This work was supported by NASA under Astrophysics Theory Program grants
10-ATP10-0171  and  NNX11AI96G.


\appendix                      
\section{Appendix: Grad-Shafranov equation(s) for the torus.}
In cylindrical coordinates, equation of motion (\ref{eq:momentConservEq}) reads

    \begin{equation}
      \rho\thinspace\left(\partial_{t}+{\bf v}\cdot\nabla\right){\bf v}=-\nabla 
P+ \mathbf{F}_{{\rm m}}-\rho\frac{G\thinspace M_{{\rm 
BH}}}{r^{2}}\thinspace{\bf \hat{r}}\mbox{.}\label{eq:EquationOfMotion1_Apdx}
    \end{equation}
    The Lorentz  force $\mathbf{\mathbf{F}_{{\rm m}}}$ reads:

    \begin{equation}
      \mathbf{F}_{{\rm m}}=\frac{{\bf J}\times{\bf B}}{c}\mbox{,}\label{eq:magneticForce1}
    \end{equation}
    where $\mathbf{B}$ is the magnetic field and the current
    density, $\mathbf{J}$ is found from:
    
    \begin{equation} 
    {\bf J}=\frac{c}{4\pi}\nabla\times{\bf
        B}\mbox{.}\label{eq:J_rotB}
    \end{equation}

\noindent
    Ideal MHD implies that $\mathbf{B}$ satisfies the
    hydromagnetic equation:

    \begin{equation}
      \partial_{t}{\bf B}=\nabla\times({\bf V\times{\bf 
B})\mbox{.}}\label{eq:rotVxB}
    \end{equation}
    It is convenient, instead of the magnetic field, $\mathbf{B}$
    to adopt a vector potential, ${\bf A}$:

    \begin{eqnarray} {\bf B} & = & \nabla\times{\bf
        A}\mbox{.}\label{eq:brotA}
    \end{eqnarray}

    Axial symmetry allows to describe all of the magnetic
    properties via a single flux function, $\Psi\sim A_{\phi}$ ,
    instead of the three components of $\mathbf{B}$. The flux
    function is defined from the following equation:

        \begin{equation}
          \Psi(R,z)=\frac{1}{2\pi}\int{\bf B}\cdot{\bf 
n}\:ds=\frac{1}{2\pi}\int{\bf \nabla\times A}\cdot{\bf 
n}\:ds=\frac{1}{2\pi}\intop_{0}^{2\pi}A_{\phi}\cdot\hat{{\bf 
\phi}\:}dl=A_{\phi}R\mbox{,}\label{eq:psiDef}
        \end{equation}
        where ${\bf n}$ is the normal vector to the equatorial
        plane, $l$ is the element of length of the circle of the
        radius $R$. From (\ref{eq:brotA}) and (\ref{eq:psiDef})
        the $r-$, and $\phi-$components of the magnetic field
        read:

        \begin{eqnarray}
          B_{r} & = & -\partial_{z}A_{\phi}=-
\frac{1}{R}\partial_{z}\Psi,\label{eq:bRFromPsi}\\
          B_{z} & = & -\frac{1}{R}\partial_{R}\Psi.\label{eq:bZFromPsi}
        \end{eqnarray}
        From (\ref{eq:bRFromPsi}),(\ref{eq:bZFromPsi}) one has:

        \begin{equation}
         \mathbf{B}=\frac{I}{R}\,\hat{\phi}+\frac{\nabla\Psi\times\hat{\phi}}{R}\mbox{,}
	\label{eq:B_vector_from_Psi_Apdx}
        \end{equation}
        where $I=RB_{\phi}$ is the poloidal current.
        We assume that the gas can be described by a polytropic
        equation of state:

        \begin{equation}
          P=K\rho^{\gamma},\label{eq:K_ro^gamma}
        \end{equation}
        where $\gamma=1+\frac{1}{n},$ where $n$ is the polytropic
        index, and $K$ is constant. The later implies that entropy
        is constant as well. Instead of pressure, $P$ it is more
        convenient, to use enthalpy, $H$, as for the equation of
        state (\ref{eq:K_ro^gamma}) it can be cast in the form:

        \begin{equation}
          H=\frac{\gamma}{\gamma-1}\frac{P}{\rho}.\label{eq:enthalpy}
        \end{equation}

        The $r-$, and $\phi-$components of the equation of motion
        (\ref{eq:EquationOfMotion1_Apdx}) transform as follows:
        \begin{eqnarray}
          \rho\,\partial_{R}(H+\phi) & = & \left(\frac{J_{\phi}}{cR}-
		\frac{I(\Psi)\,I(\Psi)'}{4\pi  R^{2}}\right)\partial_{R}\Psi+\rho\frac{v_{\phi}^{2}}{R},	   \label{eq:EqMotionReduced_R}\\
          \rho\,\partial_{z}(H+\phi) & = & \left(\frac{J_{\phi}}{cR}-
\frac{I(\Psi)\,I(\Psi)}{4\pi 
R^{2}}\right)\partial_{z}\Psi,\label{eq:EqMotionReduced_z}
        \end{eqnarray}
        where $v_{\phi}$ is the angular velocity of the gas; prime
        denotes the differentiation over $\Psi$, and we also used
        the relation $dP=\rho\,dH$ which follows from
        (\ref{eq:K_ro^gamma}) and (\ref{eq:enthalpy}).  Equations
        (\ref{eq:EqMotionReduced_R}), (\ref{eq:EqMotionReduced_z})
        can be combined into a single Grad-Shafranov equation
        which was in fact initially derived by Soloviev
        \citep{Leontovich_Vol7_Soloviev}.  The $\phi-$component of
        the current density, $J_{\phi}$ is found from
        (\ref{eq:J_rotB}):

        \begin{equation}
          J_{\phi}=\frac{c}{4\pi}(\partial_{z}B_{r}-\partial_{R}B_{z})=-
\frac{c}{4\pi}\Delta^{\star}\Psi\label{eq:J_phi}\mbox{,}
        \end{equation}
        where the ``pseudo-Laplacian'', $\Delta^{\star}$ is
        defined by the following relation:

        \begin{equation}          
	\Delta^{\star}\Psi=\partial_{z}\partial_{z}\Psi+R\partial_{R}\left(\frac{1}{R}\partial_{R}\Psi\right)\mbox{.}\label{eq:pseudoLaplace}
        \end{equation}
        From (\ref{eq:BrFunc2}) and (\ref{eq:EqMotionReduced_R})-
(\ref{eq:pseudoLaplace}) the Grad-Shafranov equation is obtained:

        \begin{equation}
          \rho\left(\nabla B_{{\rm r}}+\Omega R^{2}\nabla\Omega\right)=-
\frac{1}{4\pi R^{2}}\left(\nabla^{\star}\Psi+I\thinspace 
I\right)\nabla\Psi\mbox{.}\label{eq:GradShafranov1}
        \end{equation}
        From (\ref{eq:GradShafranov1}) now, an expression for
        $\rho$ can obtained:

        \begin{eqnarray}
          \rho & = & -\frac{1}{4\pi R^{2}}\frac{\nabla^{\star}\Psi+I\thinspace 
I'}{\partial_{R}({\rm Br}(\Psi)+\frac{\Omega^{2}R^{2}}{2})-
\Omega^{2}R}\partial_{R}\Psi=-\frac{1}{4\pi 
R^{2}}\frac{\nabla^{\star}\Psi+I\thinspace I'}{\partial_{z}({\rm 
Br}(\Psi)+\frac{\Omega^{2}R^{2}}{2})}\partial_{z}\Psi\nonumber \\
          \, & = & -\frac{1}{4\pi R^{2}}\frac{\nabla^{\star}\Psi+I\thinspace 
I'}{{\rm Br}'(\Psi)+\Omega(\Psi)\Omega'(\Psi)\thinspace 
R^{2}}\mbox{.}\label{eq:roExpression1}\\
          \nonumber 
        \end{eqnarray}
Assuming ${\rm Br} = \Omega = 0$, and assuming $\frac{dP(\Psi)}{d\Psi}=0$ one can find a Soloviev solution (\ref{eq:psiSolShort}) by direct substitution and balancing the remaining free parameters.

      \end{document}